\documentclass[%
 reprint,
superscriptaddress,
 amsmath,amssymb,
 aps
 ]{revtex4-2}
\setcitestyle{round}
\usepackage[normalem]{ulem}
\usepackage{graphicx}
\graphicspath{{im/}{../im/}}
\usepackage[dvipsnames]{xcolor}%
\usepackage{url}
\usepackage[%
    colorlinks=true,
    pdfborder={0 0 0},
    linkcolor=Blue,
    citecolor=Blue,
    urlcolor=Blue
]{hyperref}
\usepackage{orcidlink}
\usepackage{times}

\newcommand\braket[1]{\mathinner{\langle{#1}\rangle}}
\newcommand\Braket[1]{\left\langle#1\right\rangle}

\newcommand\Ket[1]{\left|#1\right\rangle}

\newcommand\diff[2]{\frac{\mathrm{d}#1}{\mathrm{d}#2}}

\begin{document}

\title{Quantum control of ion-atom collisions beyond the ultracold regime}%

\author{Maks Z. Walewski\,\orcidlink{0000-0002-9616-6421}}
\email{mz.walewski@uw.edu.pl}
\affiliation{%
Faculty of Physics, University of Warsaw, Pasteura 5, {02-093} Warsaw, Poland
}
\author{Matthew D. Frye\,\orcidlink{0000-0003-4807-2807}}%
\affiliation{%
Faculty of Physics, University of Warsaw, Pasteura 5, {02-093} Warsaw, Poland
}%
\author{Or Katz\,\orcidlink{0000-0001-7634-1993}}
\affiliation{%
School of Applied and Engineering Physics, Cornell University, Ithaca, NY 14853, USA
}%
\author{Meirav Pinkas\,\orcidlink{0000-0003-2184-0474}}
\affiliation{%
Department of Physics of Complex Systems, Weizmann Institute of Science, Rehovot 7610001, Israel
}%
\author{Roee Ozeri\,\orcidlink{0000-0001-7843-8801}}
\affiliation{%
Department of Physics of Complex Systems, Weizmann Institute of Science, Rehovot 7610001, Israel
}%
\author{Micha{\l} Tomza\,\orcidlink{0000-0003-1792-8043}}%
\email{michal.tomza@fuw.edu.pl}
\affiliation{%
Faculty of Physics, University of Warsaw, Pasteura 5, {02-093} Warsaw, Poland
}

\date{\today}%
 
\begin{abstract}
Tunable scattering resonances are crucial for controlling atomic and molecular systems.
However, their use has so far been limited to ultracold temperatures.
These conditions remain hard to achieve for most hybrid trapped ion-atom systems---a prospective platform for quantum technologies and fundamental research.
Here we measure inelastic collision probabilities for ${\text{Sr}^++\text{Rb}}$ and use them to calibrate a comprehensive theoretical model of ion-atom collisions.
Our theoretical results, compared with experimental observations, confirm that quantum interference effects persist to the multiple-partial-wave regime, leading to the pronounced state and mass dependence of the collision rates.
Using our model, we go beyond interference and identify a rich spectrum of Feshbach resonances at moderate magnetic fields with the Rb atom in its lower ($f=1$) hyperfine state, which persist at temperatures as high as $1\,\text{mK}$.
Future observation of these predicted resonances should allow precise control of the short-range dynamics in ${\text{Sr}^+}+{\text{Rb}}$ collisions under unprecedentedly warm conditions.
\end{abstract}

\maketitle

\section{Introduction}
Cooling matter near absolute zero is one of the most reliable ways to control intermolecular interactions. At ultracold temperatures, two-body collisions become dominated by a single value of orbital angular momentum $L = 0$ (\textit{s}-wave collisions), allowing collision rates to be adjusted with tunable scattering resonances. Magnetically and optically tunable Feshbach resonances have become an established tool for probing interactions and controlling chemical reactions of neutral atoms \cite{chin2010feshbach-rmp} and molecules \cite{prl2005Cs2Cs2Feshbachresonance, yang2019feshbach, wang2021feshbach, science2022feshbach-atom-molecule,nature2023feshbach-2molecules}, and only recently have been observed in ultracold ion-atom collisions \cite{nature2021ionatomresonances, thielemann2024atomionFeshbachresonances}. However, resonant control of collisions remains a challenge for most ion-atom systems, which cannot be easily cooled to the single-partial-wave regime.

In ion-atom systems, the \textit{s}-wave scattering regime is shifted down to temperatures much lower than $1\,\mu\mathrm{K}$ due to the long-range nature of their interactions~\cite{tomza2019coldhybrid}. At the same time, the oscillating electric field of the radio-frequency ion traps may heat the ion during the collision and prevent the ion-atom pair from reaching the ultracold regime \cite{cetina2012micromotion-cooling-limit, meir2016dynamics-coleed-atom-ion, pinkas2020iontrapdistribution, pinkas2023boundstates}.

At higher collision energies, the scattering state of the colliding pair is a superposition of many partial wave contributions. This often leads to averaging of quantum effects such as resonances and interference, which are therefore hard to observe. Consequently, ion-atom collisions are usually treated by the essentially classical Langevin model \cite{langevin1905, gioumousis1958langevin} at even slightly elevated temperatures. Reaching the ultracold regime has thus been considered a critical condition for observing quantum scattering effects, including Feshbach resonances, in ion-atom collisions \cite{naturephys2020ionatombuffercooling, nature2021ionatomresonances}. %

Contrary to that assumption, recent theoretical and experimental \cite{tomza2018spinimpurity, cote2018signatures, sikorsky2018phaselocking, tscherbul2023coherentcontrol} studies suggest that signatures of quantum interference can be observed in some exchange processes \cite{cote2018signatures} high above the ultracold regime due to the so-called \textit{partial-wave phase locking} effect \cite{sikorsky2018phaselocking}.
In a collision between an ion and an atom in their ${}^{2}\mathrm{S}$ electronic ground states, the relevant process is spin-exchange \cite{tomza2018spinimpurity, sikorsky2018phaselocking, katz2022quantumlogic}. 
This is driven by the difference of scattering phases acquired on scattering in the singlet and triplet electronic spin states of the system. The partial-wave phase locking effect allows the singlet-triplet phase difference to remain constant over a wide range of partial waves and collision energies. In effect, the spin-exchange cross sections for many partial waves vary in a concerted way, as if they were dominated by a single partial wave. However, this effect does not in itself suggest that collisional resonances persist to the multiple-partial-wave regime.

Here, we present a joint experimental and theoretical study of quantum effects in collisions between the Sr$^+$ ion and the Rb atom beyond the ultracold regime. We measure the probability of two types of scattering events -- a hyperfine relaxation of one neutral atom and a spin flip of a single ion -- for all initial spin projections in the $f=2$ hyperfine state of the Rb atoms. We use the results to calibrate a comprehensive theoretical model of ${\text{Sr}^++\text{Rb}}$ collisions. Employing the calibrated model, we reveal that spin-flip probabilities in the $f=1$ state of Rb can be controlled by magnetically tunable Feshbach resonances far beyond the ultracold regime. 
We predict these effects can be explored in available experimental setups up to temperatures of about $1\,\mathrm{mK}$, with as many as $15$ partial waves contributing to inelastic cross sections.

\section*{Results}
\subsection*{Measuring inelastic collisions}

\begin{figure*}[tbh]
    \centering
    \includegraphics[width=18cm]{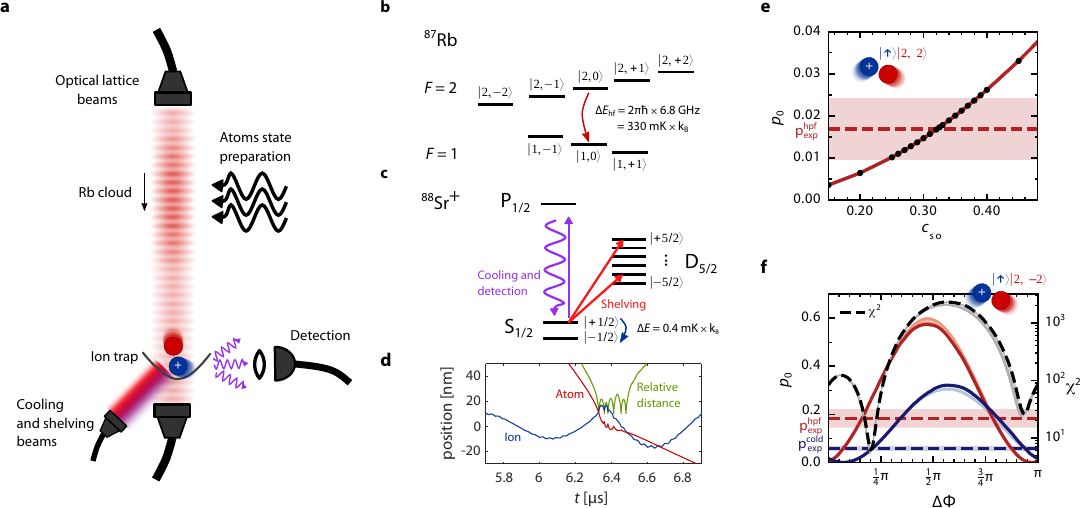} 
    \caption{\label{fig:exp-calibration}\textbf{Experimental calibration of the theoretical model of ${\text{Sr}^++\text{Rb}}$ collisions}. (\textbf{a})~Experimental setup. The atomic cloud prepared in a specific hyperfine and Zeeman state $\Ket{f,m_f}$ is optically transported to the ion's chamber.
    The ion is detected by state-selective fluorescence. 
    (\textbf{b})~The Zeeman splitting of different hyperfine manifolds of an $^{87}$Rb atom in the electronic ground state, $5^2\mathrm{S}_{1/2}$. An example of hyperfine relaxation process is denoted by the red arrow. 
    (\textbf{c})~$^{88}$Sr$^+$ energy levels scheme used for preparation and detection (repump lasers are not shown for simplicity). The cold spin-flip process is shown by the blue arrow.
    (\textbf{d})~An example of a trajectory with multiple short-range collisions due to the trap effect. Ion and atom positions are projections of motion on one of the trap axes.
    (\textbf{e})~The calculated short-range probability of the hyperfine relaxation for the spin-stretched initial state $\Ket{2,2}_\mathrm{Rb}\Ket{\uparrow}_\mathrm{Sr^+}$ as a function of the spin-orbit scaling factor $c_\mathrm{so}$. The measured value and its standard uncertainty are shown as a horizontal line with a shadow. Fit to the experimental value yields $c_\mathrm{so} = 0.32(7)$.
    (\textbf{f})~The calculated probabilities of hyperfine relaxation (red) and cold spin flip (blue) for the $\Ket{2,-2}_\mathrm{Rb}\Ket{\uparrow}_\mathrm{Sr^+}$ initial state as a function of the singlet-triplet phase difference $\Delta\Phi$. The bold lines show the results for the \textit{ab initio} value of the singlet phase $\Phi_\mathrm{s}$, and the shades behind them are for other values of $\Phi_\mathrm{s}$ from $0$ to $\pi$. The measured values are marked as dashed horizontal lines with standard uncertainties as shadows. Assuming \textit{ab initio} value of $\Phi_\mathrm{s}$, the minimum of $\chi^2=6.12$ is obtained for $\Delta\Phi=0.20(2)\pi$.}
\end{figure*}

The experimental setup is shown schematically in Fig.~\ref{fig:exp-calibration}a and described in detail in Methods; the apparatus is similar to our previous work in Refs.~\cite{ben-Shlomi2021HighEnergyResolution,katz2022quantumlogic,pinkas2023boundstates, katz2022quantum}.
Briefly, a single \textsuperscript{88}Sr$^+$ ion is trapped in a linear segmented Paul trap. It is cooled by Doppler and resolved sideband cooling, and optically pumped into its initial spin state $\Ket{\uparrow}=|\mathrm{S}_{1/2}, m_z=+1/2\rangle$.
In a separate chamber, a cloud of \textsuperscript{87}Rb atoms is loaded and cooled in a magneto-optical trap, and about $10^{6}$ atoms are loaded into a 1D optical lattice in any desired hyperfine and Zeeman state $\Ket{f,m_f}$.
The atoms are optically transported through the ion Paul trap using a travelling lattice and can collide with the ion. The average number of collisions per passage is low (approximately $0.25$) and multiple collisions are rare.

We can experimentally detect two outcomes of a scattering event. The first is a hyperfine relaxation of the atom from the upper hyperfine manifold to the lower one (red arrow in Fig.~\ref{fig:exp-calibration}b), which is measured by its impact on the ion's motion via exothermic energy release.
The second is a spin flip of the ion without changing the atom's hyperfine manifold (blue arrow in Fig.~\ref{fig:exp-calibration}c), which we can observe by directly measuring the ion's spin state.
Both processes are detected using electron-shelving and state-selective fluorescence techniques, shown in Fig.~\ref{fig:exp-calibration}c.
For \textsuperscript{84}Sr$^+$, \textsuperscript{86}Sr$^+$, and \textsuperscript{87}Sr$^+$, we use the quantum logic technique described in Ref.~\cite{katz2022quantumlogic}.

The measured probabilities of exothermic collisions in the ion trap are enhanced by the formation of temporary ion-atom bound states which are induced by the ion trapping.
These are very loosely bound and the ion-atom pair typically has a number of discrete short-range collisions before breaking up \cite{pinkas2023boundstates}; an example of this dynamics is shown in Fig.~\ref{fig:exp-calibration}d.
For each of these collisions there is a probability $p_0$ of a given event (e.g., inelastic collision), which we refer to as the \emph{short-range} probability.
We use molecular dynamics simulations to establish the relationship between $p_0$ and the measured probabilities. 
It is hard to calibrate these measurements directly to give absolute rate coefficients, so instead we normalize them to heating rates due to Langevin-type collisions.
Owing to the separation of length scales associated with the ion trap and with the chemical forces acting on the colliding pair, the short-range probabilities $p_0$ are suitable for comparison with our scattering calculations \cite{pinkas2023boundstates}.
The exact measurement and data analysis protocols are detailed in the Methods section.

\subsection*{Quantum interference effects}
The outcome of an inelastic ${\text{Sr}^++\text{Rb}}$ collision is determined by two complementary mechanisms: spin exchange and spin relaxation. Spin exchange allows the transfer of spin between the atom and the ion while keeping their total spin projection conserved.
Its effect is determined by interference between scattering on the singlet and triplet potentials, and its cross section can be approximated as~\cite{dalgarno1961spinchange}
\begin{equation}
    \label{eq:DISA}
    \sigma_\mathrm{SE} \approx \left| \Braket{\Psi_\mathrm{out}|\mathbf{\hat{s}}_\mathrm{at}\cdot\mathbf{\hat{s}}_\mathrm{ion}|\Psi_\mathrm{in}} \right|^2 \frac{4\pi}{k^2} \sum_{L=0}^\infty (2L+1) \sin^2\Delta\eta_L.
\end{equation}
Here, $\Psi_\mathrm{in}$ and $\Psi_\mathrm{out}$ are the initial and final spin states of the ${\text{Sr}^++\text{Rb}}$ pair, $\mathbf{\hat{s}}_\mathrm{at}$ and $\mathbf{\hat{s}}_\mathrm{ion}$ are the electron spin operators of the atom and the ion, $k$ is the wave number, and $\Delta\eta_L$ denotes the difference of the singlet and triplet scattering phase shifts for the given partial wave $L$.
The so-called partial-wave phase locking effect means that the singlet-triplet phase difference $\Delta\eta_L$ remains constant over a wide range of partial waves and energies \cite{cote2018signatures,sikorsky2018phaselocking,tscherbul2023coherentcontrol}, as long as the centrifugal barrier for the given $L$ is far enough below the scattering energy.
Even though the individual phases vary strongly with energy and partial wave, the conservation of this phase difference means that interference effects can persist in the spin-exchange cross section through averaging both over partial waves and over thermal energy spreads to remarkably high temperatures.
Spin relaxation, on the other hand, allows the angular momentum to be transferred between the spin and rotational degrees of freedom. In the case of ${\text{Sr}^++\text{Rb}}$ collisions, it is caused by a substantial second-order spin-orbit coupling and is perturbative.

We focus now on collisions with Rb in its upper hyperfine state $f=2$.
We measure hyperfine relaxation and cold spin-flip losses experimentally, as described above, and compare with theoretical calculations. We perform \emph{ab initio} calculations of the singlet and triplet interaction potentials, along with the second-order spin-orbit coupling, using the \textsc{Molpro} package. We then use these to perform coupled-channel calculations of the relevant scattering processes using the \textsc{molscat} program. Full details are described in Methods.

The accuracy of \textit{ab initio} electronic structure methods is insufficient for making exact predictions of the inelastic collision probabilities measured here.
Therefore, we introduce three free parameters to control our calculated interaction potentials. We allow scaling the \textit{ab initio} second-order spin-orbit coupling by a factor $c_\mathrm{so}$, and introduce the singlet and triplet phase parameters ($\Phi_\mathrm{s}$ and $\Phi_\mathrm{t}$); these are defined by the semiclassical phase integrals ${\Phi_i=\int_{R_\mathrm{cl}}^{\infty}\sqrt{-2\mu V_i (R)/\hbar^2}\,\mathrm{d}R + \pi/4}$.
The integer part of $\Phi_i/\pi$ gives the number of bound states for each potential, which is $N_s=133$ for the singlet and $N_t=271$ for the triplet potential.
We control phases $\Phi_i$ by small scaling of the short-range parts of the corresponding potential energy curves without changing the number of bound states.
Within the idea of phase locking, the difference between these semiclassical phases is a good approximation for the phase difference $\Delta\eta_L$.
According to Eq.~\eqref{eq:DISA}, only this phase difference is important for spin exchange, so we fix the singlet phase to its \textit{ab initio} value of ${\Phi_\mathrm{s}\,\mathrm{mod}\,\pi=0.045\pi}$; we have verified that it does not affect the inelastic collision probabilities in the ${f=2}$ state of Rb.

We calibrate the model by fitting the values of the spin-orbit coupling scaling $c_\mathrm{so}$ and the phase difference ${\Delta\Phi = (\Phi_\mathrm{t}-\Phi_\mathrm{s})\,\mathrm{mod}\,\pi}$.
The calibration can be performed as two separate fits for only two initial spin states of the system.
We first fit $c_\mathrm{so}$ using the probability of hyperfine relaxation from the $\Ket{2,2}_\mathrm{Rb}\Ket{\uparrow}_\mathrm{Sr^+}$ channel; we use this spin state because it is spin-stretched so cannot undergo spin-exchange and is insensitive to $\Delta\Phi$.
Here we fit to the hyperfine relaxation only because we have a better estimation of measurement errors for hyperfine relaxation than for the ion's cold spin flip.
The comparison between theory and experiment is shown in Fig.~\ref{fig:exp-calibration}e and yields $c_\mathrm{so} = 0.32(7)$.
In the same way, we fit the value of the phase difference $\Delta\Phi = 0.20(2)\pi$ to match the experimental hyperfine relaxation and the ion's cold spin-flip probabilities for the spin-exchange-dominated $\Ket{2,-2}_\mathrm{Rb}\Ket{\uparrow}_\mathrm{Sr^+}$ initial spin state of the colliding pair. Here, we need both the hyperfine relaxation and the ion's cold spin-flip probability to determine $\Delta\Phi$ unequivocally as seen from Fig.~\ref{fig:exp-calibration}f.
We neglect the spin-orbit coupling when fitting $\Delta\Phi$ to spare computational time as its effect is minuscule compared with the spin exchange for the chosen spin state.

\begin{figure}[tb]
    \centering
    \includegraphics[width=88mm]{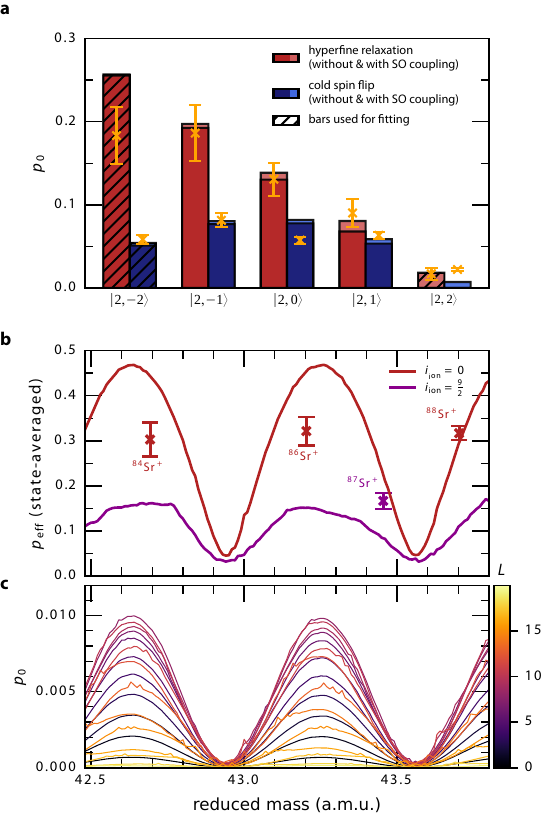} 
\caption{\label{fig:phase-locking}\textbf{Phase locking effect}. (\textbf{a})~Short-range probabilities of hyperfine relaxation (red bars) and cold spin flip of the ion (blue bars) calculated with the fitted values of $\Delta\Phi=0.2\pi$ and $c_\mathrm{so}=0.32$ for five initial spin states of Rb atoms, compared with the measured values (yellow error bars).
The \textsuperscript{88}Sr$^+$ ion was prepared in the $\Ket{\uparrow}$ spin state.
(\textbf{b})~Calculated probabilities of hyperfine relaxation averaged over the initial spin states of the ${\text{Sr}^++\text{Rb}}$ pair, plotted as a function of the reduced mass of the system, and compared with experimental values from Ref.~\cite{katz2022quantumlogic}.
Here we calculate and use the trap-enhanced probabilities $p_\mathrm{eff}$ in place of $p_0$ to enable comparison with the experiment (see Methods).
We show the results for two values of the nuclear spin of the ion, $i_\mathrm{ion} = 0$ (red line, corresponding to even isotopes) and $i_\mathrm{ion} = 9/2$ (purple line, corresponding to \textsuperscript{87}Sr).
(\textbf{c})~Partial-wave contributions to the short-range probability of the $\left|2,0\right>_\mathrm{Rb}\left|\uparrow\right>_\mathrm{Sr^+}\to\left|1,1\right>_\mathrm{Rb}\left|\downarrow\right>_\mathrm{Sr^+}$ transition, one of the possible hyperfine relaxation pathways. The largest contribution comes from $L=8$, with up to 20 partial waves involved.
}
\end{figure}

We investigate the accuracy of our calibrated model by predicting inelastic collision probabilities for other spin states, and even other isotopic combinations.
In Fig.~\ref{fig:phase-locking}a, we show in solid bars the short-range probabilities of the hyperfine relaxation and the ion's cold spin flip, calculated for the fitted values of $c_\mathrm{so}$ and $\Delta\Phi$, and compare them with the measured values for all initial spin projections in the $f_\mathrm{Rb}=2$ channel. There is a good agreement between the experimental data and the results of the scattering calculations, validating our calibrated model. We can also see the state dependence predicted by the first factor in Eq.~\eqref{eq:DISA} in both theory and experiment.

\begin{figure*}[t]
    \centering
    \includegraphics[width=18cm]{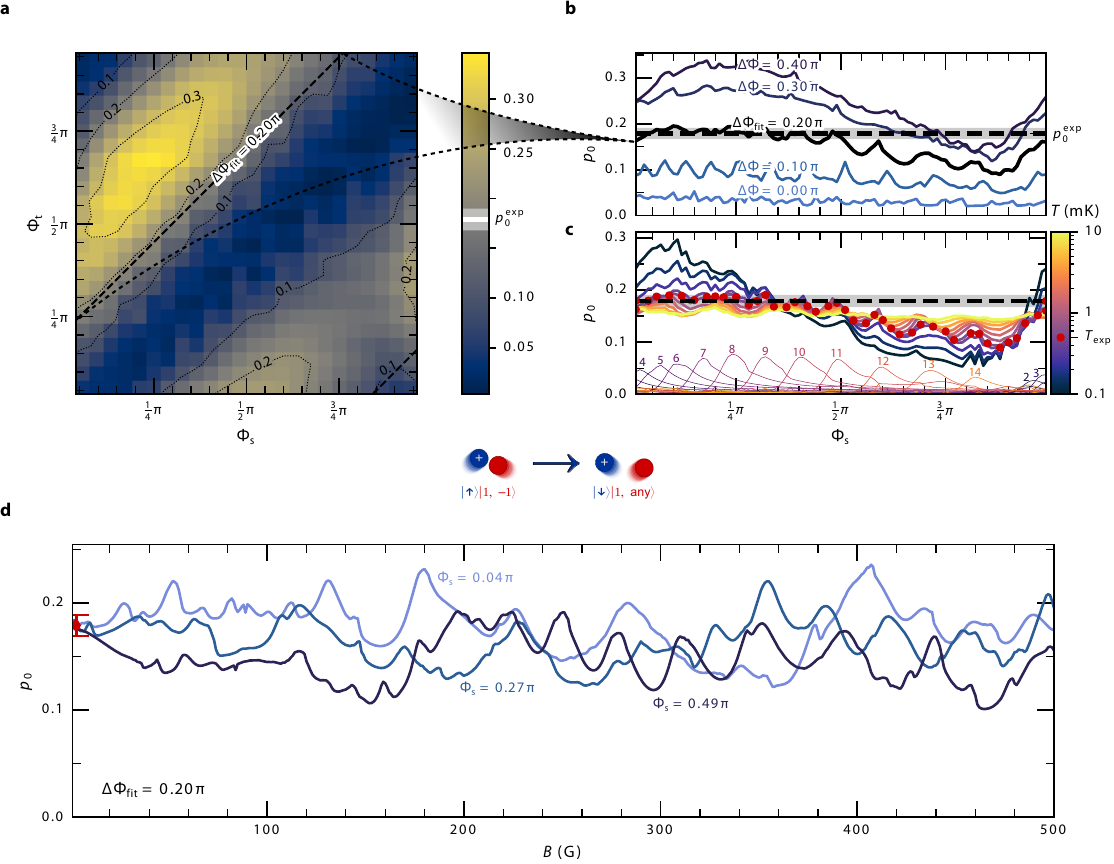}
    \caption{\label{fig:resonances}\textbf{Feshbach resonances beyond the ultracold regime}. The calculated short-range probability $p_0$ of a spin flip of the $\mathrm{{}^{88}Sr^+}$ ion, where the ion-atom pair is prepared in the $\Ket{\uparrow}_\mathrm{Sr^+}\Ket{1,-1}_\mathrm{Rb}$ initial state, compared with the experimental value $p_0^\mathrm{exp}=0.1790(99)$ measured at an external magnetic field $B=3\,\mathrm{G}$.
    (\textbf{a})~The short-range probability plotted as a function of both $\Phi_\mathrm{s}$ and $\Phi_\mathrm{t}$, calculated at $B=3\,\mathrm{G}$ from thermal averages at $T_\mathrm{exp} = 0.5\,\mathrm{mK}$, which corresponds to the experimental conditions.
    (\textbf{b})~The short-range probability calculated at $B=3\,\mathrm{G}$ and $T_\mathrm{exp} = 0.5\,\mathrm{mK}$ as a function of $\Phi_\mathrm{s}$ for a few fixed values of $\Delta\Phi$, including the fitted $\Delta\Phi_\mathrm{fit}=0.2\pi$.
    In panels~\textbf{b} and \textbf{c}, the measured value is marked as a dashed horizontal line with a shaded area marking its standard uncertainty.
    (\textbf{c})~The short-range probability calculated for the previously fixed value of $\Delta\Phi_\mathrm{fit}=0.2\pi$, plotted for a range of temperatures from $0.1$ to $10\,\mathrm{mK}$ as a function of $\Phi_\mathrm{s}$. We see the gradual loss of sensitivity to the singlet phase as the temperature rises. The probability calculated for $T_\mathrm{exp}=0.5\,\mathrm{mK}$ is indicated by red dots, and the partial-wave contributions at this temperature are labelled by the value of $L$ for the given partial wave.
    (\textbf{d})~The short-range probability calculated at $T_\mathrm{exp}=0.5\,\mathrm{mK}$ as a function of the magnetic field $B$ for 3 arbitrarily chosen values of $\Phi_\mathrm{s}$ that would match the experimental value measured at $B=3\,\mathrm{G}$. The latter is shown as a red point with an error bar representing the standard uncertainty.}
\end{figure*}

Changing the Sr$^+$ isotope changes the reduced mass and therefore the phase integrals over the potential. These vary as $\sqrt{\mu}$ for both potentials, and due to the different number of bound states in each potential the phase difference $\Delta\Phi$ also scales similarly. Following Eq.~\eqref{eq:DISA}, the result should be a sinusoidal variation in spin-exchange cross sections as a function of $\mu$.
In Fig.~\ref{fig:phase-locking}b, we present the calculated trap-enhanced hyperfine relaxation probability, averaged over the initial spin projections of the Rb atom and the Sr$^+$ ion, as a function of the reduced mass of the colliding pair, treated as a parameter in the scattering calculations, and compare it with experimental results from Ref.~\cite{katz2022quantumlogic}.
The sinusoidal shape of the curve for even Sr$^+$ isotopes is distorted by the trap effects, which result in larger enhancement of small short-range probabilities $p_0$ and lead to sharper minima compared to the rounded maxima of the $\sin^2(\mu)$ function.
Figure~\ref{fig:phase-locking}c shows how different partial waves contribute to the short-range probability of one of the possible hyperfine relaxation pathways $\left(\left|2,0\right>_\mathrm{Rb}\left|\uparrow\right>_\mathrm{Sr^+}\to\left|1,1\right>_\mathrm{Rb}\left|\downarrow\right>_\mathrm{Sr^+}\right)$, clearly showing that the oscillations due to interference remain in phase over many partial waves due to the phase-locking effect.
The periodic behaviour of the calculated probabilities as a function of both the reduced mass (Fig.~\ref{fig:phase-locking}b-c) and $\Delta\Phi$ (Fig.~\ref{fig:exp-calibration}f) is a clear indication of quantum interference far beyond the ultracold limit, persisting over many partial waves and across a broad energy range by the phase-locking mechanism.

Figure \ref{fig:phase-locking}b shows a clear interference effect, but the periodicity predicted by our calculations does not fully correspond to the values measured for different strontium isotopes.
It is rather improbable that the deviations could be explained by the errors in the \textit{ab initio} potential energy curves or the corrections to the Born-Oppenheimer approximation.
To recover the correct periodicity, we would have to scale the singlet and triplet potential energy curves by at least 20\% in opposite directions.
That is far beyond the expected errors for the electronic structure calculations as described in Methods, which should not typically exceed a few per cent.
On the other hand, the mass shifts needed to account for different periodicity are at least 4 orders of magnitude larger than the typical corrections to the Born-Oppenheimer approximation for Rb and Sr~\cite{lutz2016deviationsBOAscaling}.
The deviations of the measured values from the scattering calculations may suggest unaccounted systematic effects from the Paul trap used to store the ion. This hypothesis could be verified by weakening the trap or by investigating systems with a larger ion-to-atom mass ratio, both of which would reduce the probability of creating the bound states and possible systematic errors. However, this approach requires radical changes in the experimental sequence.

\subsection*{Quantum resonance effects}
We now turn to effects that are dependent on the individual phases $\Phi_\mathrm{s}$ and $\Phi_\mathrm{t}$, rather than just their difference $\Delta\Phi$. This requires reaching a regime in which spin-exchange according to Eq.~\eqref{eq:DISA} does not dominate. This could be at sufficiently low temperatures in which threshold and scattering-length effects dominate, but such temperatures are beyond the reach of current experiments. Instead we look at scattering in the lower hyperfine state of Rb atoms, ${f_\mathrm{Rb} = 1}$, where Eq.~\eqref{eq:DISA} does not hold because there are few outgoing channels, each with very small energy release.

We measure the ion's spin-flip probability for the atom-ion pair prepared in the $\Ket{1,-1}_\mathrm{Rb}\Ket{\uparrow}_\mathrm{Sr^+}$ spin state as described above, and perform the corresponding scattering calculations using \textsc{molscat}. In Fig.~\ref{fig:resonances}a-b, we show the calculated probability as a function of the singlet and triplet phases, $\Phi_\mathrm{s}$ and $\Phi_\mathrm{t}$, together with a few sections through the contour map for fixed values of the phase difference $\Delta\Phi$.
There is a broad dependence on the singlet phase for large phase differences, up to a factor of 2, but even for $\Delta\Phi=0.1\pi$ there are numerous smaller sharp oscillations/peaks.
For our fitted $\Delta\Phi_\mathrm{fit}=0.2\pi$ there is moderate variation, and the theory predictions agree with the experimental measurement for roughly half the range of $\Phi_\mathrm{s}\,\mathrm{mod}\,\pi$.
In Fig.~\ref{fig:resonances}c, we show how the calculated spin-flip probability varies with temperature between $0.1$ and $10\,\mathrm{mK}$. Both the broad variation and the sharp features become more pronounced at lower temperature, but persist up to several $\mathrm{mK}$.

The arguments of the phase-locking model apply only to the phase difference and not to the variation of individual phases. Features as a function of the individual phases are therefore expected to average out at these temperatures and their presence here is at first sight surprising.
Figure~\ref{fig:resonances}c also shows a breakdown of the partial-wave contributions at $T=0.5\,\mathrm{mK}$. This shows that the contributions of the partial waves peak at increasing phase $\Phi_\mathrm{s}$ in order. These cover the entire cycle of phase at this temperature, but not uniformly. The variation in height and spacing of peaks creates the broad variation in $p_0$ while individual peaks standing out above the background causes the sharper features. This is a very different behaviour than observed in Fig.~\ref{fig:phase-locking}c for $f=2$ incoming states, where all partial waves peaked together, and confirms this effect is distinct from the phase locking.

We attribute these features to the effect of Feshbach resonances originating from molecular levels of $\Ket{f=2, m_f}_\mathrm{Rb}\Ket{m_s}_\mathrm{Sr^+}$ spin states. These occur when a (quasi-)bound state is near the scattering energy and interacts with the incoming channel, and they greatly enhance inelastic scattering in their partial wave.
Due to the large binding energy (relative to their own $f=2$ thresholds) and the very strong coupling provided by the spin exchange, these resonances have large underlying widths compared to the cold temperatures of the experiment, and so can survive thermal averaging (compare Supplementary Fig.~1).
As discussed above, the positions of the peaks shift only a little between consecutive partial waves. This happens because the effective rotational constant for these states is small, both compared to the vibrational and hyperfine splittings, so only a small change in $\Phi_\mathrm{s}$ is needed to bring the next into resonance.
At $\Delta\Phi=0.1\pi$ these resonances show up as individual sharp features, but at larger $\Delta\Phi$, the increased coupling widens them so they overlap and form a single broad variation through the cycle. 
At higher temperatures, the number of resonances that contribute increases and they cover the range of $\Phi_\mathrm{s}\,\mathrm{mod}\,\pi$ more uniformly, leading to the effects becoming washed out. However, at lower temperatures, fewer resonances contribute and they are more tightly clustered, enhancing the variation.

No real experiment can vary $\Phi_\mathrm{s}$, 
but these Feshbach resonance results nonetheless suggest that resonances may exist as a function of a physically controllable parameter.
We therefore calculate the scattering as a function of the magnetic field~$B$ from 0 to 500 G.
Our calculations, presented in Fig.~\ref{fig:resonances}d, show a marked magnetic-field dependence of the spin-flip probabilities for the experimental temperature of $0.5\,\mathrm{mK}$.
As shown in Fig.~\ref{fig:resonances}c and Supplementary Fig.~2, we are right at the edge of temperatures that allow the observation of Feshbach resonances and the resonances are much more pronounced for $T\approx0.1\,\mathrm{mK}$.
The enhancement due to chosen Feshbach resonances reaches a factor of $2$, and should be observable in modern hybrid ion-atom experiments, even taking into account the intricate trap effects~\cite{pinkas2023boundstates}.
The interpretation of individual peaks is not simple but the overall pattern may act as a fingerprint, enabling us to determine $\Phi_\mathrm{s}$ and $\Phi_\mathrm{t}$ even at $T=0.5\,\mathrm{mK}$.

\section*{Discussion}
We have presented a comprehensive model of collisions between the Sr$^+$ ion and the Rb atom, capable of predicting inelastic collision probabilities in the multiple-partial-wave regime. As seen in Fig.~\ref{fig:phase-locking}a, our scattering calculations agree with the measured values for most spin states of the colliding pair, with deviations smaller than the standard uncertainty of our measurements.
The calculated hyperfine relaxation and cold spin-flip probabilities depend periodically on both $\Delta\Phi$ and the reduced mass of the system, which is a strong signature of interference persisting to temperatures many orders of magnitude higher than the ultracold regime through the phase-locking mechanism.
This allows us to determine highly sensitive short-range parameters controlling inelastic collision rates and put conditions on the interaction potentials which govern ${\text{Sr}^++\text{Rb}}$ collisions.

The magnetic Feshbach resonances predicted by our model substantially modify the spin-flip probabilities high above the ultracold regime, and should be observable in modern hybrid ion-atom systems at approachable temperatures. 
The calculated variation of the spin-flip rates is marked under the conditions of the current experiment (${T\approx0.5\,\mathrm{mK}}$), but cooling the system to ${T\approx0.1\,\mathrm{mK}}$ would result in much better resolution and contrast, still well above the \textit{s}-wave collision regime.
This will allow tuning the interactions of ion-atom pairs without the need to cool deep into the ultracold regime, opening up new avenues of control over hybrid ion-atom systems~\cite{tomza2019coldhybrid}.
Future measurements of the spectrum of resonances should allow calibrating both the singlet and triplet potential energy curves and would constitute the first observation of magnetically tunable Feshbach resonances in the multiple-partial-wave regime. 
\section*{Materials and methods}
\subsection*{Experimental apparatus}
A cloud of $^{87} $Rb atoms is loaded and cooled down in a magneto-optical trap (MOT), followed by a dark-MOT stage and polarization gradient cooling, loading a cloud of approximately $10^6$ atoms into an optical lattice formed by two counter-propagating off-resonant beams at $1064\,\mathrm{nm}$. The atoms are prepared in a specific Zeeman state in the $f=1$ or $f=2$ hyperfine manifolds by a sequence of microwave and optical pumping pulses. A \textsuperscript{88}Sr$^+$ ion is trapped in a different vacuum chamber in a Paul trap made of linear segmented blades, with secular trap frequencies $\omega = (0.49,1.21,1.44)\times2\pi~\mathrm{MHz}$, and RF frequency $\Omega_\mathrm{RF} = 26.5\times2\pi~\mathrm{MHz}$. The ion is cooled down by Doppler cooling, followed by a resolved sideband cooling scheme that cools the ion's motion down to the ground state, and pumping pulse which prepares it in the $\Ket{\uparrow}=|\mathrm{S}_{1/2}, m_z=+1/2\rangle$ state.

The atomic cloud is transported $25~\mathrm{cm}$ down to the ion's chamber by changing the relative optical frequencies of the counter-propagating optical lattice beams. The velocity of the atoms is tuned to collide with the ion at a nominal velocity of $0.24\,\mathrm{m/s}$, equivalent to an energy of about 300$\,\mu\mathrm{K}\times{}k_\mathrm{B}$ in the laboratory frame of reference. The background magnetic field during the collision is set at $2.97\,\mathrm{G}$.

To probe collisions in which the ion changes its spin and the atom remains in the same hyperfine manifold after the cloud passage through the ion trap, we apply the following sequence: two $\pi$-pulses using the shelving transition $|\mathrm{S}_{1/2},-1/2\rangle\to{}|\mathrm{D}_{5/2},-5/2\rangle$ first and $|\mathrm{S}_{1/2},-1/2\rangle\to{}|\mathrm{D}_{5/2},+3/2\rangle$ second with a $674\,\mathrm{nm}$ laser, and then detect fluorescence by driving the $\mathrm{S}_{1/2} -\mathrm{P}_{1/2}$ transition with a $422\,\mathrm{nm}$ laser. If the atom remains in the same hyperfine manifold, then the released energy is less than $1\,\mathrm{mK}$ and all transitions in the sequence are driven efficiently; a bright (dark) ion indicates a spin up (down) state. We repeat this experiment $N_\mathrm{exp}$ times and count how many events of spin down $N_\mathrm{d}$ are measured. We used $N_\mathrm{exp}=2250$ for all configurations, except for the atomic state $\Ket{f=1,m_f=-1}$ where we used $N_\mathrm{exp}=4500$.

To probe collisions in which the atom changes its hyperfine manifold, we first apply optical pumping pulses that ensure that the ion populates the $|\mathrm{S}_{1/2},-1/2\rangle$ state and then attempt to shelve the ion into the $\mathrm{D}_{5/2}$ manifold via two $\pi$ pulses, $|\mathrm{S}_{1/2},+1/2\rangle\to |\mathrm{D}_{5/2},+5/2\rangle$ and $|\mathrm{S}_{1/2},+1/2\rangle \to |\mathrm{D}_{5/2},-3/2\rangle$ \cite{katz2022quantumlogic}. Due to the internal energy released during a change of a hyperfine state into the motional degrees of freedom in the relative atom-ion frame, about ${\Delta{}E_\mathrm{hf}\approx~h\times 6.8\,\mathrm{GHz}\approx0.33\,\mathrm{K}\times{}k_\mathrm{B}}$ in the center-of-mass frame of reference, the shelving attempt fails at high efficiency, therefore maintaining the ion in the ground state. Via detection of fluorescence by driving the $\mathrm{S}_{1/2}\to\mathrm{P}_{1/2}$ transition with a $422\,\mathrm{nm}$ laser we can identify such shelving failure events, $N_\mathrm{gs}$, indicating a collision has occurred.  
We repeated both types of measurements for all channels in two different configurations, one in which the excess micromotion energy is near zero and another when it is large (about $1\,\mathrm{K}$), to enable estimation of the Langevin collision {probability}. The latter technique was first proposed in Ref.~\cite{katz2022quantum}.%

\subsection*{Analysis of experimental data}
We estimate the probability of a given scattering event from the experimental data with the aid of a numerical model detailed in Refs.~\cite{katz2022quantum,katz2022quantumlogic} to account for the various factors that affect the experimental reading. This model numerically simulates the motion of the ion in the trap including the experimental trapping parameters, micromotion effects, and the initial temperature of the ion $T$. We assume that in a given passage of the atom cloud, the ion experiences Langevin collisions drawn from a Poisson distribution with an average number of events of $\kappa_\mathrm{L}$. We consider a Langevin-type collision as an instantaneous elastic event in a random time where the ion's position is maintained but its instantaneous velocity $\textbf{v}_{\mathrm{i}}$ is updated to \cite{katz2022quantum,zipkes2011kinetics}\begin{equation} \textbf{v}_{\mathrm{i}}\rightarrow(1-r+\alpha r\mathcal{R}(\varphi_{\mathrm{L}}))(\textbf{v}_{\mathrm{i}}-\textbf{v}_{\mathrm{a}})+\textbf{v}_{\mathrm{a}},\end{equation} where $\textbf{v}_{\mathrm{a}}$ is the atom velocity which is randomly drawn from the Maxwell-Boltzmann distribution with the temperature of $10\,\mu$K. The mass ratio $r=\mu/m_{\mathrm{i}}\approx0.5$, where $\mu=m_{\mathrm{i}}m_{\mathrm{a}}/(m_{\mathrm{i}}+m_{\mathrm{a}})$ is the reduced mass, and $\mathcal{R}$ is the rotation matrix in the collision plane with the scattering angle $0\leq\phi_L\leq\pi$ drawn from the distribution described in Ref.~\cite{katz2022quantum}. The unitless factor $\alpha=\sqrt{1+2r\Delta E/(m_\mathrm{i}|{\bar{v}}_{\mathrm{ion}}|^2)}$ describes the increase of the ion's speed ${\bar{v}}_{\mathrm{ion}}\equiv r({v}_{\mathrm{ion}}-{v}_{\mathrm{atom}})$ in the center-of-mass frame \cite{katz2022quantumlogic}, gaining kinetic energy by the exothermic process of hyperfine changing collisions. We set $\Delta{}E=\Delta{}E_\mathrm{hf}$ with a probability $p_\mathrm{hf}$ per collision and $\Delta{}E=0$ otherwise. 

Owing to the trapping forces any instantaneous change of the ion's velocity leads to a change of its oscillation amplitude in the trap  $A_i$, which is updated every collision using the formalism described in Refs.~\cite{bermudez2017micromotion,katz2022quantum,katz2022quantumlogic}. Tracking this amplitude allows us to calculate the detection probability of a hot (bright) Sr$^+$ ion after a detection pulse%
\begin{equation}
    P_b=\cos^2\left(\frac{\pi}{2}\prod_{i}J_0(k_iA_i)\right),
\end{equation}
assuming a long detection pulse compared to the motional cycle. Here $k_i$ denotes the components of the shelving beam wavenumber along the modes axes, and $J_0$ is the zeroth order Bessel function. 

For each spin state we run the simulation using different sets of ($p_\mathrm{hf},\kappa_\mathrm{L}$) to match $P_b=N_{\mathrm{gs}}/N_{\mathrm{exp}}$ at both micromotion temperatures; as expected, we find that {$P_b$ is mostly determined by $\kappa_\mathrm{L}$ at the high micromotion temperature and by $p_\mathrm{hf}$ at the low micromotion temperature}. 
We repeat the simulation about $10^5$ times, ensuring convergence, and take average results. A typical value of the probability of a short-range (Langevin) collision per passage of the cloud is $\kappa_L\approx0.25$ for all channels, indicating that the probability of multiple collisions per passage of the cloud is small. The probabilities $p_{\mathrm{hf}}$ correspond to the yellow data points shown above the red bars in Fig.~\ref{fig:phase-locking}a. We estimate $T\approx0.55\,\mathrm{mK}$ for all channels to match the independent measurement of shelving failure of ions, when the atoms are prepared in the $f=1$ hyperfine manifold. This initial temperature effectively determines the collision energy of the atom-ion pair and is consistent with the scale of micromotion heating and magnetic energy release from a spin flip. Because this is an effective formalism that doesn't discern finite technical fidelity of the process from collision energy, we consider the uncertainty in $T$ as a contributor to the total error and add it in quadrature to the statistical error, as shown in Fig.~\ref{fig:phase-locking}a. For cold collisions in which the ion flips its spin, we run a similar code but match $P_\mathrm{sf}=N_{\mathrm{d}}/N_{\mathrm{exp}}$ with a unity detection efficiency.

 The probability of inelastic scattering events can be enhanced by the trap-induced ion-atom bound states as described in Ref.~\cite{pinkas2023boundstates}. A strongly exothermic collision tends to break the bound state immediately. In effect, the effective trap-enhanced probability of an exothermic collision can be calculated from its short-range probability $p_0$:
\begin{equation}
    \label{eq:methods:peff_p0}
    p_\mathrm{eff} = \sum_n \left( \sum_{k \geq n} \mathrm{PMF}(k) \right) \left(1-p_0\right)^{n-1} p_0.
\end{equation}
Here $p_\mathrm{eff}$ corresponds to either $p_\mathrm{sf}$ or $p_\mathrm{hf}$ extracted from the simulation. $\mathrm{PMF}(n)$ is the probability mass function for having exactly $n$ collisions in the bound state before its dissociation in the absence of inelastic scattering, for either scattering channel, which we estimate for our trapping configuration in Ref.~\cite{pinkas2023boundstates}.
We invert the $p_\mathrm{eff}(p_0)$ function to estimate the short-range probability $p_0$ from the measured inelastic collision probabilities in Figs.~\ref{fig:exp-calibration}e-f,~\ref{fig:phase-locking}a,~and~\ref{fig:resonances}.

The probability of hyperfine energy release for different strontium isotopes presented in Fig.~\ref{fig:phase-locking}b were measured for a spin mixture~\cite{katz2022quantumlogic}, and there is no obvious way to extract the short-range probability $p_0$ for a specific spin state or its state average from the measured data.
Instead, we convert the results of the quantum scattering calculations into the state-averaged trap-enhanced probability $p_\mathrm{eff}$ with the help of Eq.~\eqref{eq:methods:peff_p0} and compare it with the experimental values.
In the case of strontium-87 with its nuclear spin $i_{87}=9/2$, the ion's energy levels are split into two hyperfine manifolds, $f=4\text{ or }5$, which differ by $\Delta E\approx240\,\mathrm{mK}$. During the collision with a \textsuperscript{87}Rb atom, the hyperfine relaxation of the atom can be accompanied by the hyperfine excitation of the ion.
For \textsuperscript{87}Sr$^+$, we take into account the calculated probability of the hyperfine excitation exchange, but we weight the resulting contribution by a factor of $0.6$, which corresponds to the lower measurements efficiency for a smaller energy release which we estimate for this configuration.%

\subsection*{\label{sec:methods:electronic-structure-calculations}Electronic structure calculations}

For calculating the needed potential energy curves at short range~\cite{tomza2015PRA}, we make use of the \textit{ab initio} methods implemented in \textsc{Molpro} \cite{MOLPRO-WIREs, MOLPRO-JChemPhys, MOLPRO}. The singlet $(2)\,{}^{1}\Sigma^{+}$ potential energy curve was calculated at internuclear distances $R\leq50\,a_0$ using the Davidson corrected internally-contracted multireference configuration interaction method (MRCI+Q) \cite{werner1988internallyMRCI}, and the triplet $(1)\,{}^{3}\Sigma^{+}$ curve was obtained with the coupled-cluster method with single, double, and perturbative treatment of triple excitations [CCSD(T)]~\cite{bartlett2007CC, knowles1993CCopenshell}. In both cases, we use the correlation-consistent polarized quintuple-zeta basis set with weighted core and valence correlations (aug-cc-pwCV5Z)~\cite{hill2017gaussiansets}, with bond functions added for better convergence to the complete basis set limit near the potential minima.
For both Rb and Sr$^+$, the inner shell electrons are replaced by the small-core relativistic energy-consistent pseudopotentials ECP28MDF~\cite{lim2005ECPforRb, lim2006ECPforSr}.
Our singlet (triplet) potential has a depth of $800\,\mathrm{cm}^{-1}$ ($6412\,\mathrm{cm}^{-1}$) and an equilibrium distance of $14.1\,a_0$ ($9.3\,a_0$); these may be compared to previous large-core calculations~\cite{aymar2011PECs}, which gave depths of $960\,\mathrm{cm}^{-1}$ ($6544\,\mathrm{cm}^{-1}$) and equilibrium distances of $13.8\,a_0$ ($9.2\,a_0$).

The second-order spin-orbit coupling coefficient $\lambda_\mathrm{so}(R)$ was calculated using second-order perturbation theory from the non-relativistic $(1)\,{}^{3}\Sigma^{+}$ and $(1)\,{}^{3}\Pi$ electronic states of the ${\text{Sr}^++\text{Rb}}$ system as
\begin{equation}
    \lambda_\mathrm{so}(R) = \frac{2}{3} \frac{|\braket{(1){}^{3}\Sigma^{+}|\hat{H}_\mathrm{so}|(1){}^{3}\Pi}|^2}{V_{(1){}^{3}\Pi}(R)-V_{(1){}^{3}\Sigma^{+}}(R)}.
\end{equation}
Here $\braket{(1){}^{3}\Sigma^{+}|\hat{H}_\mathrm{so}|(1){}^{3}\Pi}$ is the matrix element of the spin-orbit interaction between $(1){}^{3}\Sigma^{+}$ and $(1){}^{3}\Pi$ electronic states calculated using MRCI wave functions, and $V_{(1){}^{3}\Pi}(R)$ and $V_{(1){}^{3}\Sigma^{+}}(R)$ are the associated potential energy curves~\cite{tomza2018spinimpurity}. The needed potential energy curve for the $(1){}^{3}\Pi$ state was calculated using the MRCI+Q method with the same basis set as for the $(2)\,{}^{1}\Sigma^{+}$ and $(1)\,{}^{3}\Sigma^{+}$ states. 
We show the calculated potential energy curves and the second-order spin-orbit coefficient in Supplementary Fig.~3 and give the values of the \textit{ab initio} points in Supplementary Data~1.

\subsection*{\label{sec:methods:long-range}Long-range interactions}

At large internuclear distances, the singlet $(2)\,{}^{1}\Sigma^{+}$ and triplet $(1)\,{}^{3}\Sigma^{+}$ potential energy curves attain the same long-range form $V_\mathrm{LR}(R) = V_\mathrm{ind}(R) + V_\mathrm{disp}(R)$, where
\begin{equation}
    V_\mathrm{ind}(R) = -\frac{C_4^\mathrm{(ind)}}{R^4} -\frac{C_6^\mathrm{(ind)}}{R^6} -\frac{C_8^\mathrm{(ind)}}{R^8} \ldots
\end{equation}
is the induced part of the potential energy, coming from the interaction of the charge of the ion with the induced multipole moments of the neutral atom, and
\begin{equation}
    V_\mathrm{disp}(R) = -\frac{C_6^\mathrm{(disp)}}{R^6} -\frac{C_8^\mathrm{(disp)}}{R^8} \ldots
\end{equation}
is the dispersion potential, arising from the interaction of instantaneous multipole moments of both the ion and the atom.

We calculate the induction coefficients $C_4^\mathrm{(ind)}$, $C_6^\mathrm{(ind)}$, and $C_8^\mathrm{(ind)}$ from the static dipole, quadrupole and octupole polarizabilities of the Rb atom,
$\alpha_1=319.8(5)\times 4\pi\epsilon_0\,a_0^3$~\cite{gregoire2015polarizabilitymeasurement}, $\alpha_2=6479(1)\times 4\pi\epsilon_0\,a_0^5$~\cite{wang2016polarizabilites}, $\alpha_3=2.381(44)\times10^5\times 4\pi\epsilon_0\,a_0^7$~\cite{wang2016polarizabilites}, and the charge of the ion $q=e$ as 
${C_{2n+2}^\mathrm{(ind)}=\frac{1}{2}q^2 \alpha_n/(4\pi\epsilon_0)^2}$~\cite{kaur2015LRpotentials}.
We use the dispersion coefficients ${C_6^\mathrm{(disp)}=1.845(6)\times10^3\,E_\mathrm{h}\,a_0^6}$ and $C_8^\mathrm{(disp)}=1.8321(2)\times10^5\,E_\mathrm{h}\,a_0^8$ as reported in Ref.~\cite{kaur2015LRpotentials}.
Our total $C_4$, $C_6$, and $C_8$ are calculated as $C_n=C_n^\mathrm{(ind)}+C_n^\mathrm{(disp)}$ and evaluate to ${C_4=159.9\,E_\mathrm{h}\,a_0^4}$, ${C_6=5079.0\,E_\mathrm{h}\,a_0^6}$, and ${C_8=302260\,E_\mathrm{h}\,a_0^8}$.

\subsection*{\label{seq:methods:modelling}Parametrization of the ion-atom interactions}

We adjust the potential energy curves used in the scattering calculations by tiny scaling of the short-range parts of the potential, which were calculated \textit{ab initio}.
These are then interpolated and extrapolated using a reciprocal-power reproducing kernel Hilbert space (RKHS) method of Ho and Rabitz \cite{ho1996rkhs, ho2000rkhs, soldan2000RKHS}, with the leading terms in the extrapolation constrained to the long-range coefficients given in the previous section, as described in Ref.~\cite{ho2000rkhs}.
We control the RKHS method by specifying the integer parameters $n=3$, $m=1$, $s=2$. Here $n$ dictates the number of reciprocal power terms at large internuclear separations, where the potential takes on the asymptotic form $\sum_{k=0}^{n-1} -C_{s(k+m+1)}/R^{s(k+m+1)}$~\cite{ho2000rkhs}.

To adjust the potential, we multiply the calculated \emph{ab initio} points by a scaling factor before applying the RKHS.
This allows us to efficiently and smoothly vary the short-range portion of the potential while leaving the accurately known long-range portion unaffected, and thus adjust the phase parameters $\Phi_i$ described in the main text.
On a technical level, we calculate the zero-energy scattering length $a_i$ for each potential energy curve using \textsc{molscat} and obtain the phase parameters as ${\Phi_i\,\mathrm{mod}\,\pi=\arctan(-a_i\hbar/\sqrt{2\mu{}C_4})+\pi/2}$~\cite{gribakin1993semiclassical}. The scaling factors we use differ from unity by at most $1.8\%$ for the singlet and $0.6\%$ for the triplet potential energy curves and are listed with the corresponding phases $\Phi_i$ in Supplementary Data~1.

\subsection*{Quantum scattering calculations}
To obtain the inelastic collision probabilities, we calculate the free-space inelastic and momentum-transfer rate coefficients by solving the Schr{\"o}dinger equation for the radial motion of the ion-atom pair.
The effective Hamiltonian used for the scattering calculations is the same as that described in detail in Ref.~\cite{hutson2008Cs2} for collisions of pairs of alkali-metal atoms and is written as
\begin{align}
\label{eq:methods:CC:hamiltonian}
\hat{\mathcal{H}} = &- \frac{\hbar^2}{2\mu R^2} \diff{}{R}\left(R^2 \diff{}{R}\right)+ \frac{\hat{L}^2}{2\mu R^2} + \hat{V}(R) + \nonumber \\
&+ \hat{\mathcal{H}}_\mathrm{Sr^+} + \hat{\mathcal{H}}_\mathrm{Rb} + \hat{\mathcal{H}}_\mathrm{ss+so}.
\end{align}
Here $R$ is the internuclear separation, $\mu$ the reduced mass, $\hat{L}$ is the orbital angular momentum of the relative motion of the ion and atom, $\hat{\mathcal{H}}_{\mathrm{Sr}^+}$ and $\hat{\mathcal{H}}_{\mathrm{Rb}}$ are the monomer Hamiltonians, consisting of the hyperfine and Zeeman terms, and $\hat{V}(R)$ contains the singlet $(2)^1\Sigma^+$ and triplet $(1)^3\Sigma^+$ molecular potential energy operators. The electron spin-spin dipolar and second-order spin-orbit interactions are modelled together as
\begin{align}
    \hat{\mathcal{H}}_\mathrm{ss+so} = &\left[ \frac{E_\mathrm{h}\alpha^2}{(R/a_0)^3} - c_\mathrm{so}\lambda_\mathrm{so}(R) \right] \nonumber\\
    &\times\left[ \mathbf{\hat{s}}_a \cdot \mathbf{\hat{s}}_b - 3 (\mathbf{\hat{s}}_\mathrm{a}\cdot \vec{e}_R)(\mathbf{\hat{s}}_\mathrm{b}\cdot \vec{e}_R)\right],
\end{align}
where $\alpha$ is the fine-structure constant, $\lambda_\mathrm{so}(R)$ is the \textit{ab initio} second-order spin-orbit coefficient, $c_\mathrm{so}$ is the scaling factor fixed to fit the experimental data as shown in Fig.~\ref{fig:exp-calibration}e, $\mathbf{\hat{s}}_a, \mathbf{\hat{s}}_b$ are the electronic spin operators of the atom and the ion, and $\vec{e}_R$ is a unit vector along the internuclear axis.

For most of our calculations, we expand the angular degrees of freedom of the scattering wavefunction in the uncoupled basis
\begin{equation}
    \Ket{L M_L; s_\mathrm{a} m_{s,\mathrm{a}}; i_\mathrm{a} m_{i, \mathrm{a}}; s_\mathrm{b} m_{s,\mathrm{b}}; i_\mathrm{b} m_{i, \mathrm{b}}}.
\end{equation}
Here, $L$ is the orbital angular momentum of the relative motion of the ion and the atom, $s_\mathrm{a}, s_\mathrm{b}$ and $i_\mathrm{a}, i_\mathrm{b}$ are the electronic and nuclear spins of the atom and the ion, and $ m_{s,\mathrm{a}}, m_{s,\mathrm{b}}, m_{i, \mathrm{a}}, m_{i, \mathrm{b}}$ are their respective projections on the quantization axis. Note that $i_\mathrm{b}=0$ in all our calculations with this basis, but we leave it in explicitly for generality.
We use \textsc{Molscat}~\cite{hutson2019molscat-article, mbf-github:2022} to solve the resulting coupled equations and calculate the \textit{S}-matrices for given collision energies. At small internuclear separations $R$ from $5.5\,a_0$ in the classically forbidden region to $21.0\,a_0$, where the long-range terms in the potential start to dominate, we propagate the log-derivative matrix using the diabatic modified log-derivative propagator of Manolopoulos \cite{manolopoulos1986propagator} with a fixed step size of $0.02\,a_0$. At $R=21.0\,a_0$, we switch to the log-derivative Airy propagator of Alexander and Manolopoulos \cite{alexander1984Airy, alexandermanolopoulos1987Airy} with an adaptive step size based on error estimates. The calculated $S$-matrices are then transformed to a basis built from atomic eigenfunctions,
\begin{equation}
\Ket{L M_L; (s_\mathrm{a}, i_\mathrm{a}) f_\mathrm{a} m_\mathrm{a}; (s_\mathrm{b}, i_\mathrm{b}) f_\mathrm{b} m_\mathrm{b}}.
\end{equation}
At zero field, $f_\mathrm{a}$ and $f_\mathrm{b}$ are total spins of the atom and ion; these are not strictly conserved in a magnetic field but are still nearly good quantum numbers at the low fields used here, so are useful as labels;
their respective projections $m_\mathrm{a}, m _\mathrm{b}$ are good quantum numbers.

We calculate the rate coefficients from the \textit{S}-matrix elements for 50 values of the collision energy in the centre-of-mass frame, ranging from $0.4\,\mu\mathrm{K}\times{}k_\mathrm{B}$ to $4\,\mathrm{mK}\times{}k_\mathrm{B}$ in Figs.~\ref{fig:exp-calibration}e-f~and~\ref{fig:phase-locking}, and from $0.8\,\mu\mathrm{K}\times{}k_\mathrm{B}$ to $80\,\mathrm{mK}\times{}k_\mathrm{B}$ in Fig.~\ref{fig:resonances}. We sum all $L, M_L$ contributions and thermally average the results assuming a Maxwell-Boltzmann distribution.

The momentum-transfer rate coefficients are calculated from \textit{S}-matrices as~\cite{frye2014momentumtransfer, anderlini2006thermalization}
\begin{align}
   k_\mathrm{m}(E) = &\sqrt{\frac{2E}{\mu}} \frac{\pi\hbar^2}{\mu{}E} \;\sum\limits_{L=0}^{L_\mathrm{max}-1} \Big[(4L+2) \sin^2{\delta_{L}} + \nonumber \\
   &- (4L+4) \sin{\delta_L}\sin{\delta_{L+1}}\cos\left(\delta_L - \delta_{L+1}\right)\Big],
\end{align}
where the real partial-wave phase shifts $\delta_L$ are related to the diagonal \textit{S}-matrix elements for the given spin channel by $S_L = \left|S_L\right|\,\exp(2i\delta_L)$. The above expression is valid for channels with fully elastic scattering. Here, we approximate the momentum-transfer rate coefficients for all the channels by $k_\mathrm{m}(E)$ calculated for the $\Ket{f=2,m_f=2}_\mathrm{Rb}\Ket{\uparrow}_\mathrm{Sr^+}$ spin state with neglected spin-spin and spin-orbit interactions. We calculate the short-range probabilities $p_0$ as a ratio of the thermally averaged inelastic and momentum-transfer rate coefficients. Then the effective trap-enhanced probabilities $p_\mathrm{eff}$ are calculated for Fig.~\ref{fig:phase-locking}b from $p_0$ as described above in Methods.

In the calculations of the hyperfine relaxation probability as a function of the reduced mass in Fig.~\ref{fig:phase-locking}b, we expand the scattering wavefunction in the basis of the total angular momentum of the colliding complex
\begin{equation}
\Ket{\left(L, \left(\left((s_\mathrm{a}, i_\mathrm{a}) f_\mathrm{a}, (s_\mathrm{b} i_\mathrm{b}) f_\mathrm{b}\right)\right) F_\mathrm{ab}\right) F M_F},
\end{equation}
where $F_\mathrm{ab}$ is the total spin of the atom-ion complex, $F$ is the total angular momentum of the colliding pair resulting from coupling the orbital angular momentum $L$ to the total spin $F_\mathrm{ab}$, and $M_F$ is the projection of $F$ on the quantization axis. At a non-zero magnetic field, the Zeeman terms couple states with different values of the total angular momentum $F$. In the case of \textsuperscript{87}Sr$^+$ with the nuclear spin of $i_{87}=9/2$, this inflates the time needed to solve the coupled equations beyond reasonable limits. On the other hand, at a small experimental magnetic field $B=2.97\,\mathrm{G}$, the Zeeman states of both Rb and Sr$^+$ are nearly degenerate, with spacing lower or similar to the collision energy; the energy scale for hyperfine relaxation is far larger than this (around $330\,\mathrm{mK}\times k_\textrm{B}$). We thus neglect the Zeeman interactions to perform calculations in the total angular momentum basis set. We verify the agreement between the calculations in the $\left|f,m_f\right>$ basis set at $B=2.97\,\mathrm{G}$ and the total angular momentum basis set with neglected Zeeman effect for \textsuperscript{88}Sr$^+$, where we can afford the direct comparison.

In both basis sets, we ensure numerical convergence with respect to the grid parameters, collision energies used for thermal averaging, and the number of included partial waves.
Although we used the total angular momentum basis set and neglected the Zeeman interaction for a part of the calculations, the total computational time needed for the project reached approximately $1.5\,\text{mln hours}\times\text{cpus}$, including around $0.75\,\text{mln hours}\times\text{cpus}$ for the final calculations presented in this paper.


\bibliography{RbSr_manuscript}

\providecommand{\noopsort}[1]{}\providecommand{\singleletter}[1]{#1}%
\begin{thebibliography}{53}%
\makeatletter
\providecommand \@ifxundefined [1]{%
 \@ifx{#1\undefined}
}%
\providecommand \@ifnum [1]{%
 \ifnum #1\expandafter \@firstoftwo
 \else \expandafter \@secondoftwo
 \fi
}%
\providecommand \@ifx [1]{%
 \ifx #1\expandafter \@firstoftwo
 \else \expandafter \@secondoftwo
 \fi
}%
\providecommand \natexlab [1]{#1}%
\providecommand \enquote  [1]{``#1''}%
\providecommand \bibnamefont  [1]{#1}%
\providecommand \bibfnamefont [1]{#1}%
\providecommand \citenamefont [1]{#1}%
\providecommand \href@noop [0]{\@secondoftwo}%
\providecommand \href [0]{\begingroup \@sanitize@url \@href}%
\providecommand \@href[1]{\@@startlink{#1}\@@href}%
\providecommand \@@href[1]{\endgroup#1\@@endlink}%
\providecommand \@sanitize@url [0]{\catcode `\\12\catcode `\$12\catcode `\&12\catcode `\#12\catcode `\^12\catcode `\_12\catcode `\%12\relax}%
\providecommand \@@startlink[1]{}%
\providecommand \@@endlink[0]{}%
\providecommand \url  [0]{\begingroup\@sanitize@url \@url }%
\providecommand \@url [1]{\endgroup\@href {#1}{\urlprefix }}%
\providecommand \urlprefix  [0]{URL }%
\providecommand \Eprint [0]{\href }%
\providecommand \doibase [0]{https://doi.org/}%
\providecommand \selectlanguage [0]{\@gobble}%
\providecommand \bibinfo  [0]{\@secondoftwo}%
\providecommand \bibfield  [0]{\@secondoftwo}%
\providecommand \translation [1]{[#1]}%
\providecommand \BibitemOpen [0]{}%
\providecommand \bibitemStop [0]{}%
\providecommand \bibitemNoStop [0]{.\EOS\space}%
\providecommand \EOS [0]{\spacefactor3000\relax}%
\providecommand \BibitemShut  [1]{\csname bibitem#1\endcsname}%
\let\auto@bib@innerbib\@empty
\bibitem [{\citenamefont {Chin}\ \emph {et~al.}(2010)\citenamefont {Chin}, \citenamefont {Grimm}, \citenamefont {Julienne},\ and\ \citenamefont {Tiesinga}}]{chin2010feshbach-rmp}%
  \BibitemOpen
  \bibfield  {author} {\bibinfo {author} {\bibfnamefont {C.}~\bibnamefont {Chin}}, \bibinfo {author} {\bibfnamefont {R.}~\bibnamefont {Grimm}}, \bibinfo {author} {\bibfnamefont {P.}~\bibnamefont {Julienne}},\ and\ \bibinfo {author} {\bibfnamefont {E.}~\bibnamefont {Tiesinga}},\ }\bibfield  {title} {\bibinfo {title} {{Feshbach} resonances in ultracold gases},\ }\href@noop {} {\bibfield  {journal} {\bibinfo  {journal} {Rev. Mod. Phys.}\ }\textbf {\bibinfo {volume} {82}},\ \bibinfo {pages} {1225} (\bibinfo {year} {2010})}\BibitemShut {NoStop}%
\bibitem [{\citenamefont {Chin}\ \emph {et~al.}(2005)\citenamefont {Chin}, \citenamefont {Kraemer}, \citenamefont {Mark}, \citenamefont {Herbig}, \citenamefont {Waldburger}, \citenamefont {N\"agerl},\ and\ \citenamefont {Grimm}}]{prl2005Cs2Cs2Feshbachresonance}%
  \BibitemOpen
  \bibfield  {author} {\bibinfo {author} {\bibfnamefont {C.}~\bibnamefont {Chin}}, \bibinfo {author} {\bibfnamefont {T.}~\bibnamefont {Kraemer}}, \bibinfo {author} {\bibfnamefont {M.}~\bibnamefont {Mark}}, \bibinfo {author} {\bibfnamefont {J.}~\bibnamefont {Herbig}}, \bibinfo {author} {\bibfnamefont {P.}~\bibnamefont {Waldburger}}, \bibinfo {author} {\bibfnamefont {H.-C.}\ \bibnamefont {N\"agerl}},\ and\ \bibinfo {author} {\bibfnamefont {R.}~\bibnamefont {Grimm}},\ }\bibfield  {title} {\bibinfo {title} {Observation of {Feshbach}-like resonances in collisions between ultracold molecules},\ }\href@noop {} {\bibfield  {journal} {\bibinfo  {journal} {Phys. Rev. Lett.}\ }\textbf {\bibinfo {volume} {94}},\ \bibinfo {pages} {123201} (\bibinfo {year} {2005})}\BibitemShut {NoStop}%
\bibitem [{\citenamefont {Yang}\ \emph {et~al.}(2019)\citenamefont {Yang}, \citenamefont {Zhang}, \citenamefont {Liu}, \citenamefont {Liu}, \citenamefont {Nan}, \citenamefont {Zhao},\ and\ \citenamefont {Pan}}]{yang2019feshbach}%
  \BibitemOpen
  \bibfield  {author} {\bibinfo {author} {\bibfnamefont {H.}~\bibnamefont {Yang}}, \bibinfo {author} {\bibfnamefont {D.-C.}\ \bibnamefont {Zhang}}, \bibinfo {author} {\bibfnamefont {L.}~\bibnamefont {Liu}}, \bibinfo {author} {\bibfnamefont {Y.-X.}\ \bibnamefont {Liu}}, \bibinfo {author} {\bibfnamefont {J.}~\bibnamefont {Nan}}, \bibinfo {author} {\bibfnamefont {B.}~\bibnamefont {Zhao}},\ and\ \bibinfo {author} {\bibfnamefont {J.-W.}\ \bibnamefont {Pan}},\ }\bibfield  {title} {\bibinfo {title} {Observation of magnetically tunable {F}eshbach resonances in ultracold {$^{23}$Na$^{40}$K + $^{40}$K} collisions},\ }\href@noop {} {\bibfield  {journal} {\bibinfo  {journal} {Science}\ }\textbf {\bibinfo {volume} {363}},\ \bibinfo {pages} {261} (\bibinfo {year} {2019})}\BibitemShut {NoStop}%
\bibitem [{\citenamefont {Wang}\ \emph {et~al.}(2021)\citenamefont {Wang}, \citenamefont {Frye}, \citenamefont {Su}, \citenamefont {Cao}, \citenamefont {Liu}, \citenamefont {Zhang}, \citenamefont {Yang}, \citenamefont {Hutson}, \citenamefont {Zhao}, \citenamefont {Bai},\ and\ \citenamefont {Pan}}]{wang2021feshbach}%
  \BibitemOpen
  \bibfield  {author} {\bibinfo {author} {\bibfnamefont {X.-Y.}\ \bibnamefont {Wang}}, \bibinfo {author} {\bibfnamefont {M.~D.}\ \bibnamefont {Frye}}, \bibinfo {author} {\bibfnamefont {Z.}~\bibnamefont {Su}}, \bibinfo {author} {\bibfnamefont {J.}~\bibnamefont {Cao}}, \bibinfo {author} {\bibfnamefont {L.}~\bibnamefont {Liu}}, \bibinfo {author} {\bibfnamefont {D.-C.}\ \bibnamefont {Zhang}}, \bibinfo {author} {\bibfnamefont {H.}~\bibnamefont {Yang}}, \bibinfo {author} {\bibfnamefont {J.~M.}\ \bibnamefont {Hutson}}, \bibinfo {author} {\bibfnamefont {B.}~\bibnamefont {Zhao}}, \bibinfo {author} {\bibfnamefont {C.-L.}\ \bibnamefont {Bai}},\ and\ \bibinfo {author} {\bibfnamefont {J.-W.}\ \bibnamefont {Pan}},\ }\bibfield  {title} {\bibinfo {title} {Magnetic {F}eshbach resonances in collisions of $^{23}${N}a$^{40}${K} with $^{40}${K}},\ }\href {https://doi.org/10.1088/1367-2630/ac3318} {\bibfield  {journal} {\bibinfo  {journal} {New J. Phys.}\ }\textbf {\bibinfo {volume} {23}},\ \bibinfo {pages} {115010} (\bibinfo
  {year} {2021})}\BibitemShut {NoStop}%
\bibitem [{\citenamefont {Son}\ \emph {et~al.}(2022)\citenamefont {Son}, \citenamefont {Park}, \citenamefont {Lu}, \citenamefont {Jamison}, \citenamefont {Karman},\ and\ \citenamefont {Ketterle}}]{science2022feshbach-atom-molecule}%
  \BibitemOpen
  \bibfield  {author} {\bibinfo {author} {\bibfnamefont {H.}~\bibnamefont {Son}}, \bibinfo {author} {\bibfnamefont {J.~J.}\ \bibnamefont {Park}}, \bibinfo {author} {\bibfnamefont {Y.-K.}\ \bibnamefont {Lu}}, \bibinfo {author} {\bibfnamefont {A.~O.}\ \bibnamefont {Jamison}}, \bibinfo {author} {\bibfnamefont {T.}~\bibnamefont {Karman}},\ and\ \bibinfo {author} {\bibfnamefont {W.}~\bibnamefont {Ketterle}},\ }\bibfield  {title} {\bibinfo {title} {Control of reactive collisions by quantum interference},\ }\href@noop {} {\bibfield  {journal} {\bibinfo  {journal} {Science}\ }\textbf {\bibinfo {volume} {375}},\ \bibinfo {pages} {1006} (\bibinfo {year} {2022})}\BibitemShut {NoStop}%
\bibitem [{\citenamefont {Park}\ \emph {et~al.}(2023)\citenamefont {Park}, \citenamefont {Lu}, \citenamefont {Jamison}, \citenamefont {Tscherbul},\ and\ \citenamefont {Ketterle}}]{nature2023feshbach-2molecules}%
  \BibitemOpen
  \bibfield  {author} {\bibinfo {author} {\bibfnamefont {J.~J.}\ \bibnamefont {Park}}, \bibinfo {author} {\bibfnamefont {Y.-K.}\ \bibnamefont {Lu}}, \bibinfo {author} {\bibfnamefont {A.~O.}\ \bibnamefont {Jamison}}, \bibinfo {author} {\bibfnamefont {T.~V.}\ \bibnamefont {Tscherbul}},\ and\ \bibinfo {author} {\bibfnamefont {W.}~\bibnamefont {Ketterle}},\ }\bibfield  {title} {\bibinfo {title} {A {Feshbach} resonance in collisions between triplet ground-state molecules},\ }\href@noop {} {\bibfield  {journal} {\bibinfo  {journal} {Nature}\ }\textbf {\bibinfo {volume} {614}},\ \bibinfo {pages} {54} (\bibinfo {year} {2023})}\BibitemShut {NoStop}%
\bibitem [{\citenamefont {Weckesser}\ \emph {et~al.}(2021)\citenamefont {Weckesser}, \citenamefont {Thielemann}, \citenamefont {Wiater}, \citenamefont {Wojciechowska}, \citenamefont {Karpa}, \citenamefont {Jachymski}, \citenamefont {Tomza}, \citenamefont {Walker},\ and\ \citenamefont {Schaetz}}]{nature2021ionatomresonances}%
  \BibitemOpen
  \bibfield  {author} {\bibinfo {author} {\bibfnamefont {P.}~\bibnamefont {Weckesser}}, \bibinfo {author} {\bibfnamefont {F.}~\bibnamefont {Thielemann}}, \bibinfo {author} {\bibfnamefont {D.}~\bibnamefont {Wiater}}, \bibinfo {author} {\bibfnamefont {A.}~\bibnamefont {Wojciechowska}}, \bibinfo {author} {\bibfnamefont {L.}~\bibnamefont {Karpa}}, \bibinfo {author} {\bibfnamefont {K.}~\bibnamefont {Jachymski}}, \bibinfo {author} {\bibfnamefont {M.}~\bibnamefont {Tomza}}, \bibinfo {author} {\bibfnamefont {T.}~\bibnamefont {Walker}},\ and\ \bibinfo {author} {\bibfnamefont {T.}~\bibnamefont {Schaetz}},\ }\bibfield  {title} {\bibinfo {title} {Observation of {Feshbach} resonances between a single ion and ultracold atoms},\ }\href@noop {} {\bibfield  {journal} {\bibinfo  {journal} {Nature}\ }\textbf {\bibinfo {volume} {600}},\ \bibinfo {pages} {429} (\bibinfo {year} {2021})}\BibitemShut {NoStop}%
\bibitem [{\citenamefont {Thielemann}\ \emph {et~al.}(2024)\citenamefont {Thielemann}, \citenamefont {Siemund}, \citenamefont {von Schoenfeld}, \citenamefont {Wu}, \citenamefont {Weckesser}, \citenamefont {Jachymski}, \citenamefont {Walker},\ and\ \citenamefont {Schaetz}}]{thielemann2024atomionFeshbachresonances}%
  \BibitemOpen
  \bibfield  {author} {\bibinfo {author} {\bibfnamefont {F.}~\bibnamefont {Thielemann}}, \bibinfo {author} {\bibfnamefont {J.}~\bibnamefont {Siemund}}, \bibinfo {author} {\bibfnamefont {D.}~\bibnamefont {von Schoenfeld}}, \bibinfo {author} {\bibfnamefont {W.}~\bibnamefont {Wu}}, \bibinfo {author} {\bibfnamefont {P.}~\bibnamefont {Weckesser}}, \bibinfo {author} {\bibfnamefont {K.}~\bibnamefont {Jachymski}}, \bibinfo {author} {\bibfnamefont {T.}~\bibnamefont {Walker}},\ and\ \bibinfo {author} {\bibfnamefont {T.}~\bibnamefont {Schaetz}},\ }\href@noop {} {\bibinfo {title} {Exploring atom-ion {F}eshbach resonances below the \textit{s}-wave limit}},\ \bibinfo {howpublished} {Preprint at: \url{https://doi.org/10.48550/arXiv.2406.13410}} (\bibinfo {year} {2024})\BibitemShut {NoStop}%
\bibitem [{\citenamefont {Tomza}\ \emph {et~al.}(2019)\citenamefont {Tomza}, \citenamefont {Jachymski}, \citenamefont {Gerritsma}, \citenamefont {Negretti}, \citenamefont {Calarco}, \citenamefont {Idziaszek},\ and\ \citenamefont {Julienne}}]{tomza2019coldhybrid}%
  \BibitemOpen
  \bibfield  {author} {\bibinfo {author} {\bibfnamefont {M.}~\bibnamefont {Tomza}}, \bibinfo {author} {\bibfnamefont {K.}~\bibnamefont {Jachymski}}, \bibinfo {author} {\bibfnamefont {R.}~\bibnamefont {Gerritsma}}, \bibinfo {author} {\bibfnamefont {A.}~\bibnamefont {Negretti}}, \bibinfo {author} {\bibfnamefont {T.}~\bibnamefont {Calarco}}, \bibinfo {author} {\bibfnamefont {Z.}~\bibnamefont {Idziaszek}},\ and\ \bibinfo {author} {\bibfnamefont {P.~S.}\ \bibnamefont {Julienne}},\ }\bibfield  {title} {\bibinfo {title} {Cold hybrid ion-atom systems},\ }\href@noop {} {\bibfield  {journal} {\bibinfo  {journal} {Rev. Mod. Phys.}\ }\textbf {\bibinfo {volume} {91}},\ \bibinfo {pages} {035001} (\bibinfo {year} {2019})}\BibitemShut {NoStop}%
\bibitem [{\citenamefont {Cetina}\ \emph {et~al.}(2012)\citenamefont {Cetina}, \citenamefont {Grier},\ and\ \citenamefont {Vuleti\ifmmode~\acute{c}\else \'{c}\fi{}}}]{cetina2012micromotion-cooling-limit}%
  \BibitemOpen
  \bibfield  {author} {\bibinfo {author} {\bibfnamefont {M.}~\bibnamefont {Cetina}}, \bibinfo {author} {\bibfnamefont {A.~T.}\ \bibnamefont {Grier}},\ and\ \bibinfo {author} {\bibfnamefont {V.}~\bibnamefont {Vuleti\ifmmode~\acute{c}\else \'{c}\fi{}}},\ }\bibfield  {title} {\bibinfo {title} {Micromotion-induced limit to atom-ion sympathetic cooling in {Paul} traps},\ }\href@noop {} {\bibfield  {journal} {\bibinfo  {journal} {Phys. Rev. Lett.}\ }\textbf {\bibinfo {volume} {109}},\ \bibinfo {pages} {253201} (\bibinfo {year} {2012})}\BibitemShut {NoStop}%
\bibitem [{\citenamefont {Meir}\ \emph {et~al.}(2016)\citenamefont {Meir}, \citenamefont {Sikorsky}, \citenamefont {Ben-shlomi}, \citenamefont {Akerman}, \citenamefont {Dallal},\ and\ \citenamefont {Ozeri}}]{meir2016dynamics-coleed-atom-ion}%
  \BibitemOpen
  \bibfield  {author} {\bibinfo {author} {\bibfnamefont {Z.}~\bibnamefont {Meir}}, \bibinfo {author} {\bibfnamefont {T.}~\bibnamefont {Sikorsky}}, \bibinfo {author} {\bibfnamefont {R.}~\bibnamefont {Ben-shlomi}}, \bibinfo {author} {\bibfnamefont {N.}~\bibnamefont {Akerman}}, \bibinfo {author} {\bibfnamefont {Y.}~\bibnamefont {Dallal}},\ and\ \bibinfo {author} {\bibfnamefont {R.}~\bibnamefont {Ozeri}},\ }\bibfield  {title} {\bibinfo {title} {Dynamics of a ground-state cooled ion colliding with ultracold atoms},\ }\href@noop {} {\bibfield  {journal} {\bibinfo  {journal} {Phys. Rev. Lett.}\ }\textbf {\bibinfo {volume} {117}},\ \bibinfo {pages} {243401} (\bibinfo {year} {2016})}\BibitemShut {NoStop}%
\bibitem [{\citenamefont {Pinkas}\ \emph {et~al.}(2020)\citenamefont {Pinkas}, \citenamefont {Meir}, \citenamefont {Sikorsky}, \citenamefont {Ben-Shlomi}, \citenamefont {Akerman},\ and\ \citenamefont {Ozeri}}]{pinkas2020iontrapdistribution}%
  \BibitemOpen
  \bibfield  {author} {\bibinfo {author} {\bibfnamefont {M.}~\bibnamefont {Pinkas}}, \bibinfo {author} {\bibfnamefont {Z.}~\bibnamefont {Meir}}, \bibinfo {author} {\bibfnamefont {T.}~\bibnamefont {Sikorsky}}, \bibinfo {author} {\bibfnamefont {R.}~\bibnamefont {Ben-Shlomi}}, \bibinfo {author} {\bibfnamefont {N.}~\bibnamefont {Akerman}},\ and\ \bibinfo {author} {\bibfnamefont {R.}~\bibnamefont {Ozeri}},\ }\bibfield  {title} {\bibinfo {title} {Effect of ion-trap parameters on energy distributions of ultra-cold atom{\textendash}ion mixtures},\ }\href@noop {} {\bibfield  {journal} {\bibinfo  {journal} {New J. Phys.}\ }\textbf {\bibinfo {volume} {22}},\ \bibinfo {pages} {013047} (\bibinfo {year} {2020})}\BibitemShut {NoStop}%
\bibitem [{\citenamefont {Pinkas}\ \emph {et~al.}(2023)\citenamefont {Pinkas}, \citenamefont {Katz}, \citenamefont {Wengrowicz}, \citenamefont {Akerman},\ and\ \citenamefont {Ozeri}}]{pinkas2023boundstates}%
  \BibitemOpen
  \bibfield  {author} {\bibinfo {author} {\bibfnamefont {M.}~\bibnamefont {Pinkas}}, \bibinfo {author} {\bibfnamefont {O.}~\bibnamefont {Katz}}, \bibinfo {author} {\bibfnamefont {J.}~\bibnamefont {Wengrowicz}}, \bibinfo {author} {\bibfnamefont {N.}~\bibnamefont {Akerman}},\ and\ \bibinfo {author} {\bibfnamefont {R.}~\bibnamefont {Ozeri}},\ }\bibfield  {title} {\bibinfo {title} {Trap-assisted formation of atom--ion bound states},\ }\href@noop {} {\bibfield  {journal} {\bibinfo  {journal} {Nat. Phys.}\ }\textbf {\bibinfo {volume} {19}},\ \bibinfo {pages} {1573} (\bibinfo {year} {2023})}\BibitemShut {NoStop}%
\bibitem [{\citenamefont {Langevin}(1905)}]{langevin1905}%
  \BibitemOpen
  \bibfield  {author} {\bibinfo {author} {\bibfnamefont {P.}~\bibnamefont {Langevin}},\ }\bibfield  {title} {\bibinfo {title} {A fundamental formula of kinetic theory},\ }\href@noop {} {\bibfield  {journal} {\bibinfo  {journal} {Ann. Chim. Phys.}\ }\textbf {\bibinfo {volume} {5}},\ \bibinfo {pages} {245} (\bibinfo {year} {1905})}\BibitemShut {NoStop}%
\bibitem [{\citenamefont {Gioumousis}\ and\ \citenamefont {Stevenson}(1958)}]{gioumousis1958langevin}%
  \BibitemOpen
  \bibfield  {author} {\bibinfo {author} {\bibfnamefont {G.}~\bibnamefont {Gioumousis}}\ and\ \bibinfo {author} {\bibfnamefont {D.~P.}\ \bibnamefont {Stevenson}},\ }\bibfield  {title} {\bibinfo {title} {Reactions of gaseous molecule ions with gaseous molecules. {V. T}heory},\ }\href@noop {} {\bibfield  {journal} {\bibinfo  {journal} {J. Chem. Phys.}\ }\textbf {\bibinfo {volume} {29}},\ \bibinfo {pages} {294} (\bibinfo {year} {1958})}\BibitemShut {NoStop}%
\bibitem [{\citenamefont {Feldker}\ \emph {et~al.}(2020)\citenamefont {Feldker}, \citenamefont {F\"urst}, \citenamefont {Hirzler}, \citenamefont {Ewald}, \citenamefont {Mazzanti}, \citenamefont {Wiater}, \citenamefont {Tomza},\ and\ \citenamefont {Gerritsma}}]{naturephys2020ionatombuffercooling}%
  \BibitemOpen
  \bibfield  {author} {\bibinfo {author} {\bibfnamefont {T.}~\bibnamefont {Feldker}}, \bibinfo {author} {\bibfnamefont {H.}~\bibnamefont {F\"urst}}, \bibinfo {author} {\bibfnamefont {H.}~\bibnamefont {Hirzler}}, \bibinfo {author} {\bibfnamefont {N.~V.}\ \bibnamefont {Ewald}}, \bibinfo {author} {\bibfnamefont {M.}~\bibnamefont {Mazzanti}}, \bibinfo {author} {\bibfnamefont {D.}~\bibnamefont {Wiater}}, \bibinfo {author} {\bibfnamefont {M.}~\bibnamefont {Tomza}},\ and\ \bibinfo {author} {\bibfnamefont {R.}~\bibnamefont {Gerritsma}},\ }\bibfield  {title} {\bibinfo {title} {Buffer gas cooling of a trapped ion to the quantum regime},\ }\href@noop {} {\bibfield  {journal} {\bibinfo  {journal} {Nat. Phys.}\ }\textbf {\bibinfo {volume} {16}},\ \bibinfo {pages} {413} (\bibinfo {year} {2020})}\BibitemShut {NoStop}%
\bibitem [{\citenamefont {F\"urst}\ \emph {et~al.}(2018)\citenamefont {F\"urst}, \citenamefont {Feldker}, \citenamefont {Ewald}, \citenamefont {Joger}, \citenamefont {Tomza},\ and\ \citenamefont {Gerritsma}}]{tomza2018spinimpurity}%
  \BibitemOpen
  \bibfield  {author} {\bibinfo {author} {\bibfnamefont {H.}~\bibnamefont {F\"urst}}, \bibinfo {author} {\bibfnamefont {T.}~\bibnamefont {Feldker}}, \bibinfo {author} {\bibfnamefont {N.~V.}\ \bibnamefont {Ewald}}, \bibinfo {author} {\bibfnamefont {J.}~\bibnamefont {Joger}}, \bibinfo {author} {\bibfnamefont {M.}~\bibnamefont {Tomza}},\ and\ \bibinfo {author} {\bibfnamefont {R.}~\bibnamefont {Gerritsma}},\ }\bibfield  {title} {\bibinfo {title} {Dynamics of a single ion-spin impurity in a spin-polarized atomic bath},\ }\href@noop {} {\bibfield  {journal} {\bibinfo  {journal} {Phys. Rev. A}\ }\textbf {\bibinfo {volume} {98}},\ \bibinfo {pages} {012713} (\bibinfo {year} {2018})}\BibitemShut {NoStop}%
\bibitem [{\citenamefont {C\^ot\'e}\ and\ \citenamefont {Simbotin}(2018)}]{cote2018signatures}%
  \BibitemOpen
  \bibfield  {author} {\bibinfo {author} {\bibfnamefont {R.}~\bibnamefont {C\^ot\'e}}\ and\ \bibinfo {author} {\bibfnamefont {I.}~\bibnamefont {Simbotin}},\ }\bibfield  {title} {\bibinfo {title} {Signature of the $s$-wave regime high above ultralow temperatures},\ }\href@noop {} {\bibfield  {journal} {\bibinfo  {journal} {Phys. Rev. Lett.}\ }\textbf {\bibinfo {volume} {121}},\ \bibinfo {pages} {173401} (\bibinfo {year} {2018})}\BibitemShut {NoStop}%
\bibitem [{\citenamefont {Sikorsky}\ \emph {et~al.}(2018)\citenamefont {Sikorsky}, \citenamefont {Morita}, \citenamefont {Meir}, \citenamefont {Buchachenko}, \citenamefont {Ben-shlomi}, \citenamefont {Akerman}, \citenamefont {Narevicius}, \citenamefont {Tscherbul},\ and\ \citenamefont {Ozeri}}]{sikorsky2018phaselocking}%
  \BibitemOpen
  \bibfield  {author} {\bibinfo {author} {\bibfnamefont {T.}~\bibnamefont {Sikorsky}}, \bibinfo {author} {\bibfnamefont {M.}~\bibnamefont {Morita}}, \bibinfo {author} {\bibfnamefont {Z.}~\bibnamefont {Meir}}, \bibinfo {author} {\bibfnamefont {A.~A.}\ \bibnamefont {Buchachenko}}, \bibinfo {author} {\bibfnamefont {R.}~\bibnamefont {Ben-shlomi}}, \bibinfo {author} {\bibfnamefont {N.}~\bibnamefont {Akerman}}, \bibinfo {author} {\bibfnamefont {E.}~\bibnamefont {Narevicius}}, \bibinfo {author} {\bibfnamefont {T.~V.}\ \bibnamefont {Tscherbul}},\ and\ \bibinfo {author} {\bibfnamefont {R.}~\bibnamefont {Ozeri}},\ }\bibfield  {title} {\bibinfo {title} {Phase locking between different partial waves in atom-ion spin-exchange collisions},\ }\href@noop {} {\bibfield  {journal} {\bibinfo  {journal} {Phys. Rev. Lett.}\ }\textbf {\bibinfo {volume} {121}},\ \bibinfo {pages} {173402} (\bibinfo {year} {2018})}\BibitemShut {NoStop}%
\bibitem [{\citenamefont {Devolder}\ \emph {et~al.}(2023)\citenamefont {Devolder}, \citenamefont {Brumer},\ and\ \citenamefont {Tscherbul}}]{tscherbul2023coherentcontrol}%
  \BibitemOpen
  \bibfield  {author} {\bibinfo {author} {\bibfnamefont {A.}~\bibnamefont {Devolder}}, \bibinfo {author} {\bibfnamefont {P.}~\bibnamefont {Brumer}},\ and\ \bibinfo {author} {\bibfnamefont {T.~V.}\ \bibnamefont {Tscherbul}},\ }\bibfield  {title} {\bibinfo {title} {Robust coherent control of two-body collisions beyond the ultracold regime},\ }\href@noop {} {\bibfield  {journal} {\bibinfo  {journal} {Phys. Rev. Res.}\ }\textbf {\bibinfo {volume} {5}},\ \bibinfo {pages} {L042025} (\bibinfo {year} {2023})}\BibitemShut {NoStop}%
\bibitem [{\citenamefont {Katz}\ \emph {et~al.}(2022{\natexlab{a}})\citenamefont {Katz}, \citenamefont {Pinkas}, \citenamefont {Akerman},\ and\ \citenamefont {Ozeri}}]{katz2022quantumlogic}%
  \BibitemOpen
  \bibfield  {author} {\bibinfo {author} {\bibfnamefont {O.}~\bibnamefont {Katz}}, \bibinfo {author} {\bibfnamefont {M.}~\bibnamefont {Pinkas}}, \bibinfo {author} {\bibfnamefont {N.}~\bibnamefont {Akerman}},\ and\ \bibinfo {author} {\bibfnamefont {R.}~\bibnamefont {Ozeri}},\ }\bibfield  {title} {\bibinfo {title} {Quantum logic detection of collisions between single atom–ion pairs},\ }\href@noop {} {\bibfield  {journal} {\bibinfo  {journal} {Nat. Phys.}\ }\textbf {\bibinfo {volume} {18}},\ \bibinfo {pages} {533–537} (\bibinfo {year} {2022}{\natexlab{a}})}\BibitemShut {NoStop}%
\bibitem [{\citenamefont {Ben-Shlomi}\ \emph {et~al.}(2021)\citenamefont {Ben-Shlomi}, \citenamefont {Pinkas}, \citenamefont {Meir}, \citenamefont {Sikorsky}, \citenamefont {Katz}, \citenamefont {Akerman},\ and\ \citenamefont {Ozeri}}]{ben-Shlomi2021HighEnergyResolution}%
  \BibitemOpen
  \bibfield  {author} {\bibinfo {author} {\bibfnamefont {R.}~\bibnamefont {Ben-Shlomi}}, \bibinfo {author} {\bibfnamefont {M.}~\bibnamefont {Pinkas}}, \bibinfo {author} {\bibfnamefont {Z.}~\bibnamefont {Meir}}, \bibinfo {author} {\bibfnamefont {T.}~\bibnamefont {Sikorsky}}, \bibinfo {author} {\bibfnamefont {O.}~\bibnamefont {Katz}}, \bibinfo {author} {\bibfnamefont {N.}~\bibnamefont {Akerman}},\ and\ \bibinfo {author} {\bibfnamefont {R.}~\bibnamefont {Ozeri}},\ }\bibfield  {title} {\bibinfo {title} {High-energy-resolution measurements of an ultracold-atom-ion collisional cross section},\ }\href@noop {} {\bibfield  {journal} {\bibinfo  {journal} {Phys. Rev. A}\ }\textbf {\bibinfo {volume} {103}},\ \bibinfo {pages} {1} (\bibinfo {year} {2021})}\BibitemShut {NoStop}%
\bibitem [{\citenamefont {Katz}\ \emph {et~al.}(2022{\natexlab{b}})\citenamefont {Katz}, \citenamefont {Pinkas}, \citenamefont {Akerman},\ and\ \citenamefont {Ozeri}}]{katz2022quantum}%
  \BibitemOpen
  \bibfield  {author} {\bibinfo {author} {\bibfnamefont {O.}~\bibnamefont {Katz}}, \bibinfo {author} {\bibfnamefont {M.}~\bibnamefont {Pinkas}}, \bibinfo {author} {\bibfnamefont {N.}~\bibnamefont {Akerman}},\ and\ \bibinfo {author} {\bibfnamefont {R.}~\bibnamefont {Ozeri}},\ }\href@noop {} {\bibinfo {title} {Quantum suppression of cold reactions far from the quantum regime}},\ \bibinfo {howpublished} {Preprint at: \url{https://doi.org/10.48550/arXiv.2208.07725}} (\bibinfo {year} {2022}{\natexlab{b}})\BibitemShut {NoStop}%
\bibitem [{\citenamefont {Dalgarno}(1961)}]{dalgarno1961spinchange}%
  \BibitemOpen
  \bibfield  {author} {\bibinfo {author} {\bibfnamefont {A.}~\bibnamefont {Dalgarno}},\ }\bibfield  {title} {\bibinfo {title} {Spin-change cross-sections},\ }\href@noop {} {\bibfield  {journal} {\bibinfo  {journal} {Proc. R. Soc. Lond.}\ }\textbf {\bibinfo {volume} {262}},\ \bibinfo {pages} {132} (\bibinfo {year} {1961})}\BibitemShut {NoStop}%
\bibitem [{\citenamefont {Lutz}\ and\ \citenamefont {Hutson}(2016)}]{lutz2016deviationsBOAscaling}%
  \BibitemOpen
  \bibfield  {author} {\bibinfo {author} {\bibfnamefont {J.~J.}\ \bibnamefont {Lutz}}\ and\ \bibinfo {author} {\bibfnamefont {J.~M.}\ \bibnamefont {Hutson}},\ }\bibfield  {title} {\bibinfo {title} {Deviations from {Born}-{Oppenheimer} mass scaling in spectroscopy and ultracold molecular physics},\ }\href@noop {} {\bibfield  {journal} {\bibinfo  {journal} {J. Mol. Spectrosc.}\ }\textbf {\bibinfo {volume} {330}},\ \bibinfo {pages} {43} (\bibinfo {year} {2016})}\BibitemShut {NoStop}%
\bibitem [{\citenamefont {Zipkes}\ \emph {et~al.}(2011)\citenamefont {Zipkes}, \citenamefont {Ratschbacher}, \citenamefont {Sias},\ and\ \citenamefont {K{\"o}hl}}]{zipkes2011kinetics}%
  \BibitemOpen
  \bibfield  {author} {\bibinfo {author} {\bibfnamefont {C.}~\bibnamefont {Zipkes}}, \bibinfo {author} {\bibfnamefont {L.}~\bibnamefont {Ratschbacher}}, \bibinfo {author} {\bibfnamefont {C.}~\bibnamefont {Sias}},\ and\ \bibinfo {author} {\bibfnamefont {M.}~\bibnamefont {K{\"o}hl}},\ }\bibfield  {title} {\bibinfo {title} {Kinetics of a single trapped ion in an ultracold buffer gas},\ }\href@noop {} {\bibfield  {journal} {\bibinfo  {journal} {New J. Phys.}\ }\textbf {\bibinfo {volume} {13}},\ \bibinfo {pages} {053020} (\bibinfo {year} {2011})}\BibitemShut {NoStop}%
\bibitem [{\citenamefont {Bermudez}\ \emph {et~al.}(2017)\citenamefont {Bermudez}, \citenamefont {Schindler}, \citenamefont {Monz}, \citenamefont {Blatt},\ and\ \citenamefont {Müller}}]{bermudez2017micromotion}%
  \BibitemOpen
  \bibfield  {author} {\bibinfo {author} {\bibfnamefont {A.}~\bibnamefont {Bermudez}}, \bibinfo {author} {\bibfnamefont {P.}~\bibnamefont {Schindler}}, \bibinfo {author} {\bibfnamefont {T.}~\bibnamefont {Monz}}, \bibinfo {author} {\bibfnamefont {R.}~\bibnamefont {Blatt}},\ and\ \bibinfo {author} {\bibfnamefont {M.}~\bibnamefont {Müller}},\ }\bibfield  {title} {\bibinfo {title} {Micromotion-enabled improvement of quantum logic gates with trapped ions},\ }\href@noop {} {\bibfield  {journal} {\bibinfo  {journal} {New J. Phys.}\ }\textbf {\bibinfo {volume} {19}},\ \bibinfo {pages} {113038} (\bibinfo {year} {2017})}\BibitemShut {NoStop}%
\bibitem [{\citenamefont {Tomza}\ \emph {et~al.}(2015)\citenamefont {Tomza}, \citenamefont {Koch},\ and\ \citenamefont {Moszynski}}]{tomza2015PRA}%
  \BibitemOpen
  \bibfield  {author} {\bibinfo {author} {\bibfnamefont {M.}~\bibnamefont {Tomza}}, \bibinfo {author} {\bibfnamefont {C.~P.}\ \bibnamefont {Koch}},\ and\ \bibinfo {author} {\bibfnamefont {R.}~\bibnamefont {Moszynski}},\ }\bibfield  {title} {\bibinfo {title} {Cold interactions between an {Y}b$^+$ ion and a {L}i atom: {P}rospects for sympathetic cooling, radiative association, and {F}eshbach resonances},\ }\href@noop {} {\bibfield  {journal} {\bibinfo  {journal} {Phys. Rev. A}\ }\textbf {\bibinfo {volume} {91}},\ \bibinfo {pages} {042706} (\bibinfo {year} {2015})}\BibitemShut {NoStop}%
\bibitem [{\citenamefont {Werner}\ \emph {et~al.}(2012)\citenamefont {Werner}, \citenamefont {Knowles}, \citenamefont {Knizia}, \citenamefont {Manby},\ and\ \citenamefont {Schütz}}]{MOLPRO-WIREs}%
  \BibitemOpen
  \bibfield  {author} {\bibinfo {author} {\bibfnamefont {H.-J.}\ \bibnamefont {Werner}}, \bibinfo {author} {\bibfnamefont {P.~J.}\ \bibnamefont {Knowles}}, \bibinfo {author} {\bibfnamefont {G.}~\bibnamefont {Knizia}}, \bibinfo {author} {\bibfnamefont {F.~R.}\ \bibnamefont {Manby}},\ and\ \bibinfo {author} {\bibfnamefont {M.}~\bibnamefont {Schütz}},\ }\bibfield  {title} {\bibinfo {title} {\textsc{Molpro}: a general-purpose quantum chemistry program package},\ }\href@noop {} {\bibfield  {journal} {\bibinfo  {journal} {WIREs Comput. Mol. Sci.}\ }\textbf {\bibinfo {volume} {2}},\ \bibinfo {pages} {242} (\bibinfo {year} {2012})}\BibitemShut {NoStop}%
\bibitem [{\citenamefont {Werner}\ \emph {et~al.}(2020)\citenamefont {Werner}, \citenamefont {Knowles}, \citenamefont {Manby}, \citenamefont {Black}, \citenamefont {Doll}, \citenamefont {Heßelmann}, \citenamefont {Kats}, \citenamefont {Köhn}, \citenamefont {Korona}, \citenamefont {Kreplin}, \citenamefont {Ma}, \citenamefont {Miller}, \citenamefont {Mitrushchenkov}, \citenamefont {Peterson}, \citenamefont {Polyak}, \citenamefont {Rauhut},\ and\ \citenamefont {Sibaev}}]{MOLPRO-JChemPhys}%
  \BibitemOpen
  \bibfield  {author} {\bibinfo {author} {\bibfnamefont {H.-J.}\ \bibnamefont {Werner}}, \bibinfo {author} {\bibfnamefont {P.~J.}\ \bibnamefont {Knowles}}, \bibinfo {author} {\bibfnamefont {F.~R.}\ \bibnamefont {Manby}}, \bibinfo {author} {\bibfnamefont {J.~A.}\ \bibnamefont {Black}}, \bibinfo {author} {\bibfnamefont {K.}~\bibnamefont {Doll}}, \bibinfo {author} {\bibfnamefont {A.}~\bibnamefont {Heßelmann}}, \bibinfo {author} {\bibfnamefont {D.}~\bibnamefont {Kats}}, \bibinfo {author} {\bibfnamefont {A.}~\bibnamefont {Köhn}}, \bibinfo {author} {\bibfnamefont {T.}~\bibnamefont {Korona}}, \bibinfo {author} {\bibfnamefont {D.~A.}\ \bibnamefont {Kreplin}}, \bibinfo {author} {\bibfnamefont {Q.}~\bibnamefont {Ma}}, \bibinfo {author} {\bibfnamefont {I.}~\bibnamefont {Miller}, \bibfnamefont {Thomas~F.}}, \bibinfo {author} {\bibfnamefont {A.}~\bibnamefont {Mitrushchenkov}}, \bibinfo {author} {\bibfnamefont {K.~A.}\ \bibnamefont {Peterson}}, \bibinfo {author} {\bibfnamefont {I.}~\bibnamefont {Polyak}}, \bibinfo
  {author} {\bibfnamefont {G.}~\bibnamefont {Rauhut}},\ and\ \bibinfo {author} {\bibfnamefont {M.}~\bibnamefont {Sibaev}},\ }\bibfield  {title} {\bibinfo {title} {The \textsc{Molpro} quantum chemistry package},\ }\href@noop {} {\bibfield  {journal} {\bibinfo  {journal} {J. Chem. Phys.}\ }\textbf {\bibinfo {volume} {152}},\ \bibinfo {pages} {144107} (\bibinfo {year} {2020})}\BibitemShut {NoStop}%
\bibitem [{\citenamefont {Werner}\ \emph {et~al.}()\citenamefont {Werner}, \citenamefont {Knowles}, \citenamefont {Knizia}, \citenamefont {Manby}, \citenamefont {{Sch\"{u}tz}}, \citenamefont {Celani}, \citenamefont {Gy\"orffy}, \citenamefont {Kats}, \citenamefont {Korona}, \citenamefont {Lindh}, \citenamefont {Mitrushenkov}, \citenamefont {Rauhut}, \citenamefont {Shamasundar}, \citenamefont {Adler}, \citenamefont {Amos}, \citenamefont {Bennie}, \citenamefont {Bernhardsson}, \citenamefont {Berning}, \citenamefont {Cooper}, \citenamefont {Deegan}, \citenamefont {Dobbyn}, \citenamefont {Eckert}, \citenamefont {Goll}, \citenamefont {Hampel}, \citenamefont {Hesselmann}, \citenamefont {Hetzer}, \citenamefont {Hrenar}, \citenamefont {Jansen}, \citenamefont {K\"oppl}, \citenamefont {Lee}, \citenamefont {Liu}, \citenamefont {Lloyd}, \citenamefont {Ma}, \citenamefont {Mata}, \citenamefont {May}, \citenamefont {McNicholas}, \citenamefont {Meyer}, \citenamefont {{Miller III}}, \citenamefont {Mura}, \citenamefont
  {Nicklass}, \citenamefont {O'Neill}, \citenamefont {Palmieri}, \citenamefont {Peng}, \citenamefont {Pfl\"uger}, \citenamefont {Pitzer}, \citenamefont {Reiher}, \citenamefont {Shiozaki}, \citenamefont {Stoll}, \citenamefont {Stone}, \citenamefont {Tarroni}, \citenamefont {Thorsteinsson}, \citenamefont {Wang},\ and\ \citenamefont {Wel{Born}}}]{MOLPRO}%
  \BibitemOpen
  \bibfield  {author} {\bibinfo {author} {\bibfnamefont {H.-J.}\ \bibnamefont {Werner}}, \bibinfo {author} {\bibfnamefont {P.~J.}\ \bibnamefont {Knowles}}, \bibinfo {author} {\bibfnamefont {G.}~\bibnamefont {Knizia}}, \bibinfo {author} {\bibfnamefont {F.~R.}\ \bibnamefont {Manby}}, \bibinfo {author} {\bibfnamefont {M.}~\bibnamefont {{Sch\"{u}tz}}}, \bibinfo {author} {\bibfnamefont {P.}~\bibnamefont {Celani}}, \bibinfo {author} {\bibfnamefont {W.}~\bibnamefont {Gy\"orffy}}, \bibinfo {author} {\bibfnamefont {D.}~\bibnamefont {Kats}}, \bibinfo {author} {\bibfnamefont {T.}~\bibnamefont {Korona}}, \bibinfo {author} {\bibfnamefont {R.}~\bibnamefont {Lindh}}, \bibinfo {author} {\bibfnamefont {A.}~\bibnamefont {Mitrushenkov}}, \bibinfo {author} {\bibfnamefont {G.}~\bibnamefont {Rauhut}}, \bibinfo {author} {\bibfnamefont {K.~R.}\ \bibnamefont {Shamasundar}}, \bibinfo {author} {\bibfnamefont {T.~B.}\ \bibnamefont {Adler}}, \bibinfo {author} {\bibfnamefont {R.~D.}\ \bibnamefont {Amos}}, \bibinfo {author} {\bibfnamefont
  {S.~J.}\ \bibnamefont {Bennie}}, \bibinfo {author} {\bibfnamefont {A.}~\bibnamefont {Bernhardsson}}, \bibinfo {author} {\bibfnamefont {A.}~\bibnamefont {Berning}}, \bibinfo {author} {\bibfnamefont {D.~L.}\ \bibnamefont {Cooper}}, \bibinfo {author} {\bibfnamefont {M.~J.~O.}\ \bibnamefont {Deegan}}, \bibinfo {author} {\bibfnamefont {A.~J.}\ \bibnamefont {Dobbyn}}, \bibinfo {author} {\bibfnamefont {F.}~\bibnamefont {Eckert}}, \bibinfo {author} {\bibfnamefont {E.}~\bibnamefont {Goll}}, \bibinfo {author} {\bibfnamefont {C.}~\bibnamefont {Hampel}}, \bibinfo {author} {\bibfnamefont {A.}~\bibnamefont {Hesselmann}}, \bibinfo {author} {\bibfnamefont {G.}~\bibnamefont {Hetzer}}, \bibinfo {author} {\bibfnamefont {T.}~\bibnamefont {Hrenar}}, \bibinfo {author} {\bibfnamefont {G.}~\bibnamefont {Jansen}}, \bibinfo {author} {\bibfnamefont {C.}~\bibnamefont {K\"oppl}}, \bibinfo {author} {\bibfnamefont {S.~J.~R.}\ \bibnamefont {Lee}}, \bibinfo {author} {\bibfnamefont {Y.}~\bibnamefont {Liu}}, \bibinfo {author} {\bibfnamefont
  {A.~W.}\ \bibnamefont {Lloyd}}, \bibinfo {author} {\bibfnamefont {Q.}~\bibnamefont {Ma}}, \bibinfo {author} {\bibfnamefont {R.~A.}\ \bibnamefont {Mata}}, \bibinfo {author} {\bibfnamefont {A.~J.}\ \bibnamefont {May}}, \bibinfo {author} {\bibfnamefont {S.~J.}\ \bibnamefont {McNicholas}}, \bibinfo {author} {\bibfnamefont {W.}~\bibnamefont {Meyer}}, \bibinfo {author} {\bibfnamefont {T.~F.}\ \bibnamefont {{Miller III}}}, \bibinfo {author} {\bibfnamefont {M.~E.}\ \bibnamefont {Mura}}, \bibinfo {author} {\bibfnamefont {A.}~\bibnamefont {Nicklass}}, \bibinfo {author} {\bibfnamefont {D.~P.}\ \bibnamefont {O'Neill}}, \bibinfo {author} {\bibfnamefont {P.}~\bibnamefont {Palmieri}}, \bibinfo {author} {\bibfnamefont {D.}~\bibnamefont {Peng}}, \bibinfo {author} {\bibfnamefont {K.}~\bibnamefont {Pfl\"uger}}, \bibinfo {author} {\bibfnamefont {R.}~\bibnamefont {Pitzer}}, \bibinfo {author} {\bibfnamefont {M.}~\bibnamefont {Reiher}}, \bibinfo {author} {\bibfnamefont {T.}~\bibnamefont {Shiozaki}}, \bibinfo {author}
  {\bibfnamefont {H.}~\bibnamefont {Stoll}}, \bibinfo {author} {\bibfnamefont {A.~J.}\ \bibnamefont {Stone}}, \bibinfo {author} {\bibfnamefont {R.}~\bibnamefont {Tarroni}}, \bibinfo {author} {\bibfnamefont {T.}~\bibnamefont {Thorsteinsson}}, \bibinfo {author} {\bibfnamefont {M.}~\bibnamefont {Wang}},\ and\ \bibinfo {author} {\bibfnamefont {M.}~\bibnamefont {Wel{Born}}},\ }\href@noop {} {\bibinfo {title} {Molpro, version 2019, a package of ab initio programs}},\ \bibinfo {note} {see \url{https://www.molpro.net}}\BibitemShut {NoStop}%
\bibitem [{\citenamefont {Werner}\ and\ \citenamefont {Knowles}(1988)}]{werner1988internallyMRCI}%
  \BibitemOpen
  \bibfield  {author} {\bibinfo {author} {\bibfnamefont {H.}~\bibnamefont {Werner}}\ and\ \bibinfo {author} {\bibfnamefont {P.~J.}\ \bibnamefont {Knowles}},\ }\bibfield  {title} {\bibinfo {title} {An efficient internally contracted multiconfiguration–reference configuration interaction method},\ }\href@noop {} {\bibfield  {journal} {\bibinfo  {journal} {J. Chem. Phys.}\ }\textbf {\bibinfo {volume} {89}},\ \bibinfo {pages} {5803} (\bibinfo {year} {1988})}\BibitemShut {NoStop}%
\bibitem [{\citenamefont {Bartlett}\ and\ \citenamefont {Musia\l{}}(2007)}]{bartlett2007CC}%
  \BibitemOpen
  \bibfield  {author} {\bibinfo {author} {\bibfnamefont {R.~J.}\ \bibnamefont {Bartlett}}\ and\ \bibinfo {author} {\bibfnamefont {M.}~\bibnamefont {Musia\l{}}},\ }\bibfield  {title} {\bibinfo {title} {Coupled-cluster theory in quantum chemistry},\ }\href@noop {} {\bibfield  {journal} {\bibinfo  {journal} {Rev. Mod. Phys.}\ }\textbf {\bibinfo {volume} {79}},\ \bibinfo {pages} {291} (\bibinfo {year} {2007})}\BibitemShut {NoStop}%
\bibitem [{\citenamefont {Knowles}\ \emph {et~al.}(1993)\citenamefont {Knowles}, \citenamefont {Hampel},\ and\ \citenamefont {Werner}}]{knowles1993CCopenshell}%
  \BibitemOpen
  \bibfield  {author} {\bibinfo {author} {\bibfnamefont {P.~J.}\ \bibnamefont {Knowles}}, \bibinfo {author} {\bibfnamefont {C.}~\bibnamefont {Hampel}},\ and\ \bibinfo {author} {\bibfnamefont {H.}~\bibnamefont {Werner}},\ }\bibfield  {title} {\bibinfo {title} {Coupled cluster theory for high spin, open shell reference wave functions},\ }\href@noop {} {\bibfield  {journal} {\bibinfo  {journal} {J. Chem. Phys.}\ }\textbf {\bibinfo {volume} {99}},\ \bibinfo {pages} {5219} (\bibinfo {year} {1993})}\BibitemShut {NoStop}%
\bibitem [{\citenamefont {Hill}\ and\ \citenamefont {Peterson}(2017)}]{hill2017gaussiansets}%
  \BibitemOpen
  \bibfield  {author} {\bibinfo {author} {\bibfnamefont {J.~G.}\ \bibnamefont {Hill}}\ and\ \bibinfo {author} {\bibfnamefont {K.~A.}\ \bibnamefont {Peterson}},\ }\bibfield  {title} {\bibinfo {title} {Gaussian basis sets for use in correlated molecular calculations. {XI}. {P}seudopotential-based and all-electron relativistic basis sets for alkali metal ({K}--{Fr}) and alkaline earth ({Ca}--{Ra}) elements},\ }\href@noop {} {\bibfield  {journal} {\bibinfo  {journal} {J. Chem. Phys.}\ }\textbf {\bibinfo {volume} {147}},\ \bibinfo {pages} {244106} (\bibinfo {year} {2017})}\BibitemShut {NoStop}%
\bibitem [{\citenamefont {Lim}\ \emph {et~al.}(2005)\citenamefont {Lim}, \citenamefont {Schwerdtfeger}, \citenamefont {Metz},\ and\ \citenamefont {Stoll}}]{lim2005ECPforRb}%
  \BibitemOpen
  \bibfield  {author} {\bibinfo {author} {\bibfnamefont {I.~S.}\ \bibnamefont {Lim}}, \bibinfo {author} {\bibfnamefont {P.}~\bibnamefont {Schwerdtfeger}}, \bibinfo {author} {\bibfnamefont {B.}~\bibnamefont {Metz}},\ and\ \bibinfo {author} {\bibfnamefont {H.}~\bibnamefont {Stoll}},\ }\bibfield  {title} {\bibinfo {title} {All-electron and relativistic pseudopotential studies for the group 1 element polarizabilities from {K} to element 119},\ }\href@noop {} {\bibfield  {journal} {\bibinfo  {journal} {J. Chem. Phys.}\ }\textbf {\bibinfo {volume} {122}},\ \bibinfo {pages} {104103} (\bibinfo {year} {2005})}\BibitemShut {NoStop}%
\bibitem [{\citenamefont {Lim}\ \emph {et~al.}(2006)\citenamefont {Lim}, \citenamefont {Stoll},\ and\ \citenamefont {Schwerdtfeger}}]{lim2006ECPforSr}%
  \BibitemOpen
  \bibfield  {author} {\bibinfo {author} {\bibfnamefont {I.~S.}\ \bibnamefont {Lim}}, \bibinfo {author} {\bibfnamefont {H.}~\bibnamefont {Stoll}},\ and\ \bibinfo {author} {\bibfnamefont {P.}~\bibnamefont {Schwerdtfeger}},\ }\bibfield  {title} {\bibinfo {title} {Relativistic small-core energy-consistent pseudopotentials for the alkaline-earth elements from {Ca} to {Ra}},\ }\href@noop {} {\bibfield  {journal} {\bibinfo  {journal} {J. Chem. Phys.}\ }\textbf {\bibinfo {volume} {124}},\ \bibinfo {pages} {034107} (\bibinfo {year} {2006})}\BibitemShut {NoStop}%
\bibitem [{\citenamefont {Aymar}\ \emph {et~al.}(2011)\citenamefont {Aymar}, \citenamefont {Guérout},\ and\ \citenamefont {Dulieu}}]{aymar2011PECs}%
  \BibitemOpen
  \bibfield  {author} {\bibinfo {author} {\bibfnamefont {M.}~\bibnamefont {Aymar}}, \bibinfo {author} {\bibfnamefont {R.}~\bibnamefont {Guérout}},\ and\ \bibinfo {author} {\bibfnamefont {O.}~\bibnamefont {Dulieu}},\ }\bibfield  {title} {\bibinfo {title} {Structure of the alkali-metal-atom + strontium molecular ions: Towards photoassociation and formation of cold molecular ions},\ }\href@noop {} {\bibfield  {journal} {\bibinfo  {journal} {J. Chem. Phys.}\ }\textbf {\bibinfo {volume} {135}},\ \bibinfo {pages} {064305} (\bibinfo {year} {2011})}\BibitemShut {NoStop}%
\bibitem [{\citenamefont {Gregoire}\ \emph {et~al.}(2015)\citenamefont {Gregoire}, \citenamefont {Hromada}, \citenamefont {Holmgren}, \citenamefont {Trubko},\ and\ \citenamefont {Cronin}}]{gregoire2015polarizabilitymeasurement}%
  \BibitemOpen
  \bibfield  {author} {\bibinfo {author} {\bibfnamefont {M.~D.}\ \bibnamefont {Gregoire}}, \bibinfo {author} {\bibfnamefont {I.}~\bibnamefont {Hromada}}, \bibinfo {author} {\bibfnamefont {W.~F.}\ \bibnamefont {Holmgren}}, \bibinfo {author} {\bibfnamefont {R.}~\bibnamefont {Trubko}},\ and\ \bibinfo {author} {\bibfnamefont {A.~D.}\ \bibnamefont {Cronin}},\ }\bibfield  {title} {\bibinfo {title} {Measurements of the ground-state polarizabilities of {Cs}, {Rb}, and {K} using atom interferometry},\ }\href@noop {} {\bibfield  {journal} {\bibinfo  {journal} {Phys. Rev. A}\ }\textbf {\bibinfo {volume} {92}},\ \bibinfo {pages} {052513} (\bibinfo {year} {2015})}\BibitemShut {NoStop}%
\bibitem [{\citenamefont {Wang}\ \emph {et~al.}(2016)\citenamefont {Wang}, \citenamefont {Jiang}, \citenamefont {Xie}, \citenamefont {Zhang},\ and\ \citenamefont {Dong}}]{wang2016polarizabilites}%
  \BibitemOpen
  \bibfield  {author} {\bibinfo {author} {\bibfnamefont {X.}~\bibnamefont {Wang}}, \bibinfo {author} {\bibfnamefont {J.}~\bibnamefont {Jiang}}, \bibinfo {author} {\bibfnamefont {L.-Y.}\ \bibnamefont {Xie}}, \bibinfo {author} {\bibfnamefont {D.-H.}\ \bibnamefont {Zhang}},\ and\ \bibinfo {author} {\bibfnamefont {C.-Z.}\ \bibnamefont {Dong}},\ }\bibfield  {title} {\bibinfo {title} {Polarizabilities and tune-out wavelengths of the hyperfine ground states of $^{87,85}\mathrm{Rb}$},\ }\href@noop {} {\bibfield  {journal} {\bibinfo  {journal} {Phys. Rev. A}\ }\textbf {\bibinfo {volume} {94}},\ \bibinfo {pages} {052510} (\bibinfo {year} {2016})}\BibitemShut {NoStop}%
\bibitem [{\citenamefont {Kaur}\ \emph {et~al.}(2015)\citenamefont {Kaur}, \citenamefont {Nandy}, \citenamefont {Arora},\ and\ \citenamefont {Sahoo}}]{kaur2015LRpotentials}%
  \BibitemOpen
  \bibfield  {author} {\bibinfo {author} {\bibfnamefont {J.}~\bibnamefont {Kaur}}, \bibinfo {author} {\bibfnamefont {D.~K.}\ \bibnamefont {Nandy}}, \bibinfo {author} {\bibfnamefont {B.}~\bibnamefont {Arora}},\ and\ \bibinfo {author} {\bibfnamefont {B.~K.}\ \bibnamefont {Sahoo}},\ }\bibfield  {title} {\bibinfo {title} {Properties of alkali-metal atoms and alkaline-earth-metal ions for an accurate estimate of their long-range interactions},\ }\href@noop {} {\bibfield  {journal} {\bibinfo  {journal} {Phys. Rev. A}\ }\textbf {\bibinfo {volume} {91}},\ \bibinfo {pages} {012705} (\bibinfo {year} {2015})}\BibitemShut {NoStop}%
\bibitem [{\citenamefont {Ho}\ and\ \citenamefont {Rabitz}(1996)}]{ho1996rkhs}%
  \BibitemOpen
  \bibfield  {author} {\bibinfo {author} {\bibfnamefont {T.}~\bibnamefont {Ho}}\ and\ \bibinfo {author} {\bibfnamefont {H.}~\bibnamefont {Rabitz}},\ }\bibfield  {title} {\bibinfo {title} {A general method for constructing multidimensional molecular potential energy surfaces from {\it ab initio} calculations},\ }\href@noop {} {\bibfield  {journal} {\bibinfo  {journal} {J. Chem. Phys.}\ }\textbf {\bibinfo {volume} {104}},\ \bibinfo {pages} {2584} (\bibinfo {year} {1996})}\BibitemShut {NoStop}%
\bibitem [{\citenamefont {Ho}\ and\ \citenamefont {Rabitz}(2000)}]{ho2000rkhs}%
  \BibitemOpen
  \bibfield  {author} {\bibinfo {author} {\bibfnamefont {T.~S.}\ \bibnamefont {Ho}}\ and\ \bibinfo {author} {\bibfnamefont {H.}~\bibnamefont {Rabitz}},\ }\bibfield  {title} {\bibinfo {title} {Proper construction of ab initio global potential surfaces with accurate long-range interactions},\ }\href@noop {} {\bibfield  {journal} {\bibinfo  {journal} {J. Chem. Phys.}\ }\textbf {\bibinfo {volume} {113}},\ \bibinfo {pages} {3960} (\bibinfo {year} {2000})}\BibitemShut {NoStop}%
\bibitem [{\citenamefont {Soldán}\ and\ \citenamefont {Hutson}(2000)}]{soldan2000RKHS}%
  \BibitemOpen
  \bibfield  {author} {\bibinfo {author} {\bibfnamefont {P.}~\bibnamefont {Soldán}}\ and\ \bibinfo {author} {\bibfnamefont {J.~M.}\ \bibnamefont {Hutson}},\ }\bibfield  {title} {\bibinfo {title} {On the long-range and short-range behavior of potentials from reproducing kernel {Hilbert} space interpolation},\ }\href@noop {} {\bibfield  {journal} {\bibinfo  {journal} {J. Chem. Phys.}\ }\textbf {\bibinfo {volume} {112}},\ \bibinfo {pages} {4415} (\bibinfo {year} {2000})}\BibitemShut {NoStop}%
\bibitem [{\citenamefont {Gribakin}\ and\ \citenamefont {Flambaum}(1993)}]{gribakin1993semiclassical}%
  \BibitemOpen
  \bibfield  {author} {\bibinfo {author} {\bibfnamefont {G.~F.}\ \bibnamefont {Gribakin}}\ and\ \bibinfo {author} {\bibfnamefont {V.~V.}\ \bibnamefont {Flambaum}},\ }\bibfield  {title} {\bibinfo {title} {Calculation of the scattering length in atomic collisions using the semiclassical approximation},\ }\href@noop {} {\bibfield  {journal} {\bibinfo  {journal} {Phys. Rev. A}\ }\textbf {\bibinfo {volume} {48}},\ \bibinfo {pages} {546} (\bibinfo {year} {1993})}\BibitemShut {NoStop}%
\bibitem [{\citenamefont {Hutson}\ \emph {et~al.}(2008)\citenamefont {Hutson}, \citenamefont {Tiesinga},\ and\ \citenamefont {Julienne}}]{hutson2008Cs2}%
  \BibitemOpen
  \bibfield  {author} {\bibinfo {author} {\bibfnamefont {J.~M.}\ \bibnamefont {Hutson}}, \bibinfo {author} {\bibfnamefont {E.}~\bibnamefont {Tiesinga}},\ and\ \bibinfo {author} {\bibfnamefont {P.~S.}\ \bibnamefont {Julienne}},\ }\bibfield  {title} {\bibinfo {title} {Avoided crossings between bound states of ultracold cesium dimers},\ }\href@noop {} {\bibfield  {journal} {\bibinfo  {journal} {Phys. Rev. A}\ }\textbf {\bibinfo {volume} {78}},\ \bibinfo {pages} {052703} (\bibinfo {year} {2008})}\BibitemShut {NoStop}%
\bibitem [{\citenamefont {Hutson}\ and\ \citenamefont {{Le Sueur}}(2019)}]{hutson2019molscat-article}%
  \BibitemOpen
  \bibfield  {author} {\bibinfo {author} {\bibfnamefont {J.~M.}\ \bibnamefont {Hutson}}\ and\ \bibinfo {author} {\bibfnamefont {C.~R.}\ \bibnamefont {{Le Sueur}}},\ }\bibfield  {title} {\bibinfo {title} {molscat: A program for non-reactive quantum scattering calculations on atomic and molecular collisions},\ }\href@noop {} {\bibfield  {journal} {\bibinfo  {journal} {Comput. Phys. Commun.}\ }\textbf {\bibinfo {volume} {241}},\ \bibinfo {pages} {9} (\bibinfo {year} {2019})}\BibitemShut {NoStop}%
\bibitem [{\citenamefont {Hutson}\ and\ \citenamefont {Le~Sueur}(2020)}]{mbf-github:2022}%
  \BibitemOpen
  \bibfield  {author} {\bibinfo {author} {\bibfnamefont {J.~M.}\ \bibnamefont {Hutson}}\ and\ \bibinfo {author} {\bibfnamefont {C.~R.}\ \bibnamefont {Le~Sueur}},\ }\href@noop {} {\bibinfo {title} {{\sc molscat}, {\sc bound} and {\sc field}, version 2020.01}},\ \bibinfo {howpublished} {\url{https://github.com/molscat/molscat/tree/35b3f597c4e8548285ed82276e48bf8b8d890a41}} (\bibinfo {year} {2020})\BibitemShut {NoStop}%
\bibitem [{\citenamefont {Manolopoulos}(1986)}]{manolopoulos1986propagator}%
  \BibitemOpen
  \bibfield  {author} {\bibinfo {author} {\bibfnamefont {D.~E.}\ \bibnamefont {Manolopoulos}},\ }\bibfield  {title} {\bibinfo {title} {An improved log derivative method for inelastic scattering},\ }\href@noop {} {\bibfield  {journal} {\bibinfo  {journal} {J. Chem. Phys.}\ }\textbf {\bibinfo {volume} {85}},\ \bibinfo {pages} {6425} (\bibinfo {year} {1986})}\BibitemShut {NoStop}%
\bibitem [{\citenamefont {Alexander}(1984)}]{alexander1984Airy}%
  \BibitemOpen
  \bibfield  {author} {\bibinfo {author} {\bibfnamefont {M.~H.}\ \bibnamefont {Alexander}},\ }\bibfield  {title} {\bibinfo {title} {Hybrid quantum scattering algorithms for long‐range potentials},\ }\href@noop {} {\bibfield  {journal} {\bibinfo  {journal} {J. Chem. Phys.}\ }\textbf {\bibinfo {volume} {81}},\ \bibinfo {pages} {4510} (\bibinfo {year} {1984})}\BibitemShut {NoStop}%
\bibitem [{\citenamefont {Alexander}\ and\ \citenamefont {Manolopoulos}(1987)}]{alexandermanolopoulos1987Airy}%
  \BibitemOpen
  \bibfield  {author} {\bibinfo {author} {\bibfnamefont {M.~H.}\ \bibnamefont {Alexander}}\ and\ \bibinfo {author} {\bibfnamefont {D.~E.}\ \bibnamefont {Manolopoulos}},\ }\bibfield  {title} {\bibinfo {title} {A stable linear reference potential algorithm for solution of the quantum close‐coupled equations in molecular scattering theory},\ }\href@noop {} {\bibfield  {journal} {\bibinfo  {journal} {J. Chem. Phys.}\ }\textbf {\bibinfo {volume} {86}},\ \bibinfo {pages} {2044} (\bibinfo {year} {1987})}\BibitemShut {NoStop}%
\bibitem [{\citenamefont {Frye}\ and\ \citenamefont {Hutson}(2014)}]{frye2014momentumtransfer}%
  \BibitemOpen
  \bibfield  {author} {\bibinfo {author} {\bibfnamefont {M.~D.}\ \bibnamefont {Frye}}\ and\ \bibinfo {author} {\bibfnamefont {J.~M.}\ \bibnamefont {Hutson}},\ }\bibfield  {title} {\bibinfo {title} {Collision cross sections for the thermalization of cold gases},\ }\href@noop {} {\bibfield  {journal} {\bibinfo  {journal} {Phys. Rev. A}\ }\textbf {\bibinfo {volume} {89}},\ \bibinfo {pages} {052705} (\bibinfo {year} {2014})}\BibitemShut {NoStop}%
\bibitem [{\citenamefont {Anderlini}\ and\ \citenamefont {Gu\'ery-Odelin}(2006)}]{anderlini2006thermalization}%
  \BibitemOpen
  \bibfield  {author} {\bibinfo {author} {\bibfnamefont {M.}~\bibnamefont {Anderlini}}\ and\ \bibinfo {author} {\bibfnamefont {D.}~\bibnamefont {Gu\'ery-Odelin}},\ }\bibfield  {title} {\bibinfo {title} {Thermalization in mixtures of ultracold gases},\ }\href@noop {} {\bibfield  {journal} {\bibinfo  {journal} {Phys. Rev. A}\ }\textbf {\bibinfo {volume} {73}},\ \bibinfo {pages} {032706} (\bibinfo {year} {2006})}\BibitemShut {NoStop}%
\end{thebibliography}%


\vspace{1em}
\textbf{Funding}: We gratefully acknowledge the Israeli Science Foundation (grant no.~1385/19), the European Union (ERC, QuantMol, 101042989), and the National Science Center, Poland (grant no.~2020/38/E/ST2/00564) for the financial support and Poland's high-performance computing infrastructure PLGrid (HPC Centers: ACK Cyfronet AGH) for providing computer facilities and support (computational grant no.~PLG/2023/016115).

\textbf{Author contributions}:
MZW\ performed the scattering calculations supervised by MT\ and MDF.
MT\ performed the electronic structure calculations.
OK\ and MP\ operated the experimental setup, collected data, analyzed it numerically and performed molecular dynamics simulations involving the trap.
MZW, MDF, MP, and OK wrote the manuscript.
RO\ and MT\ supervised the project, reviewed and edited the manuscript.
All authors worked on the interpretation of the data and contributed to the final manuscript.

\textbf{Competing interests}: The authors declare they have no competing interests.

\textbf{Data and materials availability}:
All data needed to evaluate the conclusions in the paper are present in the paper and/or the Supplementary Materials.
In particular, source data for the figures, the probability mass function, and the calculated interaction potentials are provided with the paper in the data file.

\onecolumngrid
\newpage
\begin{center}
	\textbf{\large Supplementary Materials}
\end{center}
\vspace{1cm}
\setcounter{figure}{0} 

\begin{figure}[hb!]
    \centering
    \includegraphics[width=15cm]{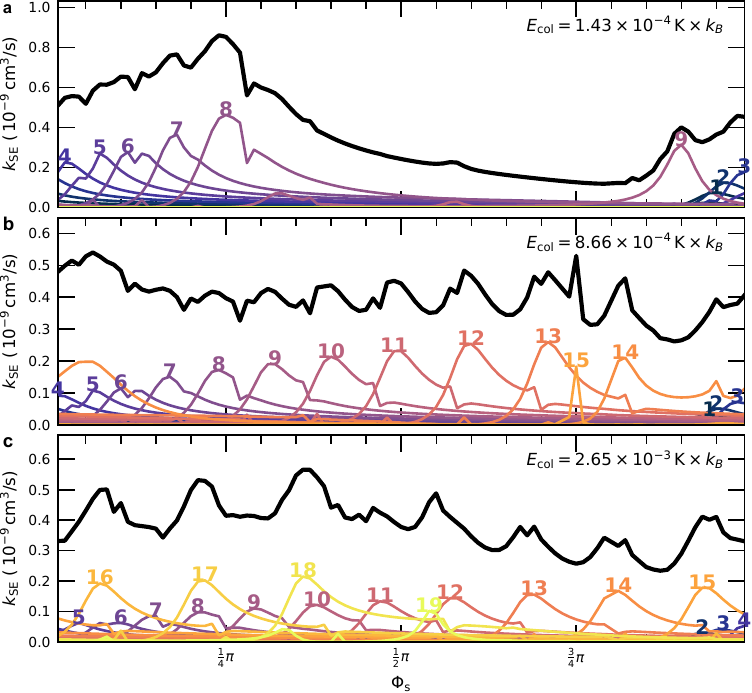} 
    \renewcommand{\figurename}{SUPPLEMENTARY FIG.}
    \caption{\label{fig:supplementary:resonances-vs-Phis}\textbf{Partial-wave contributions to the energy-resolved rate coefficients}. Spin-exchange rate coefficients ($\left|1,-1\right>\left|\uparrow\right>\to\left|1,0\right>\left|\downarrow\right>$) at fixed collision energies are plotted as a function of the singlet phase $\Phi_\mathrm{s}$ for all relevant partial waves. The phase difference is fixed to $\Delta\Phi=0.2\pi$ coming from the fit to the experimental data. At low collision energies (\textbf{a}), the relatively small number of partial waves introduces a large variation of the total inelastic rate coefficient (black line). At intermediate energies (\textbf{b}), most typical for $T_\mathrm{exp}\approx{}0.5\,\mathrm{mK}$, the variation is suppressed as more partial waves cover the full range of $\Phi_\mathrm{s}$ from $0$ to $\pi$. At higher collision energies (\textbf{c}), any possible enhancement comes from the different number of peaks from individual partial waves contributing to the total rate for different values of $\Phi_\mathrm{s}$. Notably, positions of the peaks for individual partial waves remain stable over a broad energy range in the millikelvin regime.}
\end{figure}
\clearpage
\begin{figure}
    \centering
    \includegraphics[width=18cm]{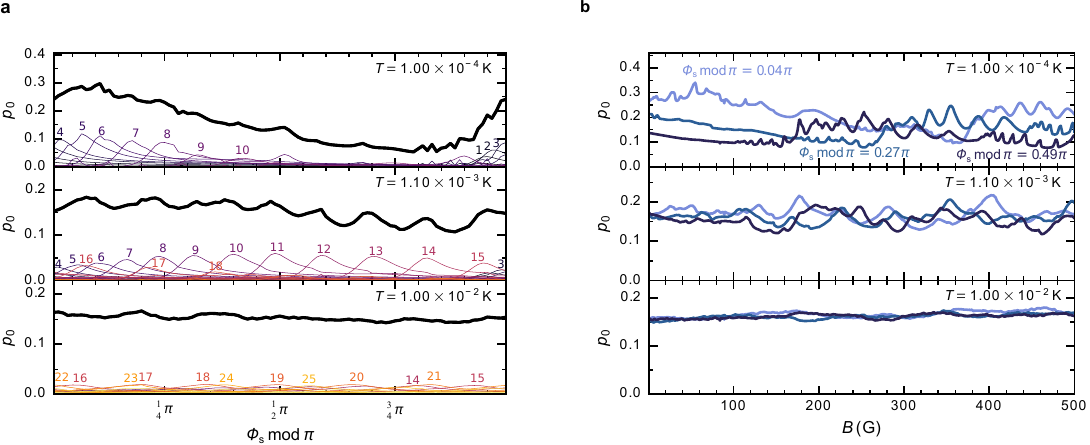}
    \renewcommand{\figurename}{SUPPLEMENTARY FIG.}
    \caption{\label{fig:supplementary:resonances}\textbf{Feshbach resonances in the multiple-partial-wave regime}.
    Calculated short-range probability $p_0$ of a spin flip of the $\mathrm{{}^{88}Sr^+}$ ion, prepared in the $\Ket{\uparrow}$ spin state, after a collision with an \textsuperscript{87}Rb atom in the $\Ket{1,-1}$ spin state, assuming the fitted value of phase difference $\Delta\Phi_\mathrm{fit}=0.2\pi$ at three different temperatures ($0.1$, $1$, and $10\,\mathrm{mK}$).
    (\textbf{a})~The probability calculated at $B=2.97\,\mathrm{G}$ as a function of $\Phi_\mathrm{s}$ is indicated by the black solid line at each temperature, and the partial-wave contributions are labelled by the value of $L$.
    (\textbf{b})~The probability calculated as a function of magnetic field for three different values of $\Phi_\mathrm{s}\,\mathrm{mod}\,\pi = 0.04\pi, 0.27\pi, 0.49\pi$ at each temperature.}
\end{figure}

\begin{figure}
    \centering
    \includegraphics[width=8.6cm]{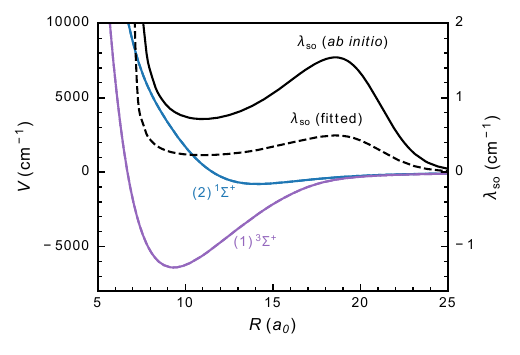} 
    \renewcommand{\figurename}{SUPPLEMENTARY FIG.}
    \caption{\label{fig:supplementary:potentials}\textbf{Interaction potentials}.
    Calculated (\textit{ab initio}) singlet $(2)\,{}^{1}\Sigma^{+}$ and triplet $(1)\,{}^{3}\Sigma^{+}$ potential energy curves are indicated by the blue and purple solid lines, respectively. Our \textit{ab initio} second-order spin-orbit coefficient $\lambda_\mathrm{so}$ is drawn as the black solid line, and $\lambda_\mathrm{so}$ multiplied by the fitted value of $c_\mathrm{so}=0.32$ is indicated by the black dashed line.
    }
\end{figure}

\setcounter{figure}{0} 
\begin{figure}
    \centering
    \renewcommand{\figurename}{SUPPLEMENTARY DATA}
    \caption{\label{fig:supplementary:dataS1}\textbf{Source data for main text figures and the interaction potentials.} The source data for the appropriate subplots in Figs.~1-3 is provided as separate sheets in the .xlsx file. In the final sheet, our \textit{ab initio} $(2)\,{}^{1}\Sigma^{+}$ and $(1)\,{}^{3}\Sigma^{+}$ potential energy curves are provided in the numerical form for the internuclear distance $R<50\,a_0$ together with the van der Waals coefficients and parameters needed for the RKHS interpolation and extrapolation. We also provide our calculated second-order spin-orbit coupling $\lambda_\mathrm{so}(R)$ and the short-range scaling factors applied to the ab initio potential energy curves to obtain the given values of semiclassical phases $\Phi_\mathrm{s}, \Phi_\mathrm{t}$.}
\end{figure}

\end{document}